\documentclass[12pt]{article}
\usepackage{mathrsfs}
\usepackage{amsfonts}
\usepackage{caption2}
\usepackage{natbib}
\usepackage{amsmath}
\usepackage{amsthm}
\usepackage{amssymb}
\usepackage{color}
\usepackage{subfigure}
\usepackage{multirow}

\usepackage[top=0.98in,bottom=0.98in,left=0.85in,right=0.85in]{geometry}
\usepackage{graphicx}

\usepackage{amssymb}
\def\be{\begin{equation}}
\def\ee{\end{equation}}
\def\bea{\begin{eqnarray}}
\def\eea{\end{eqnarray}}
\def\bd{\begin{displaymath}}
\def\ed{\end{displaymath}}
\def\bda{\begin{eqnarray*}}
\def\eda{\end{eqnarray*}}

\def\nn{\nonumber}

\def\ha1{\hat \beta_1}

\def\bd{{\mathbb D}}
\def\bb{{\mathbb B}}
\def\bi{{\mathbb I}}
\def\ba{{\mathbb A}}
\def\bw{{\mathbb W}}
\def\bx{{\mathbb X}}
\def\by{{\mathbb Y}}

\def\bmu{{\mu}}
\def\bPsi{{\Psi}}

\def\I1{{\bf I}_{\beta^{(1)}}}
\def\T{{ \mathrm{\scriptscriptstyle T} }}
\def\s1{{ \mathrm{\scriptscriptstyle (1)} }}
\def\ss2{{ \mathrm{\scriptscriptstyle (2)} }}
\def\sx{{ \mathrm{\scriptscriptstyle X} }}
\def\sa{{ \mathrm{\scriptscriptstyle A}_n }}
\def\sb{{ \mathrm{\scriptscriptstyle U}_n }}
\def\su{{ \mathrm{\scriptscriptstyle U}_n }}

\def\sg{{ \mathrm{\scriptscriptstyle G} }}

\def\bsc{\begin{scriptsize}}
\def\esc{\end{scriptsize}}

\newtheorem{tm}{Theorem}
\newtheorem{la}{Lemma}[section]
\newtheorem{cy}{Corollary}
\newtheorem{pn}{Proposition}

\theoremstyle{definition}
\newtheorem{as}{Assumption}[section]

\newcommand{\ov}{$\hfill \square$}

\begin{document}
\begin{Large}
\title{ Tests for High Dimensional Generalized Linear Models \footnote{{\bf The research paper was stole by someone last November and illegally submitted to arXiv by a person named gong
zi jiang nan. We have asked arXiv to withdraw the unfinished paper
and it was removed last December. We have collected enough evidences
to identify the person and Peking University has begun to
investigate the plagiarizer.}}}
\author{\normalsize  Song Xi Chen and Bin Guo\\
\normalsize{Peking University and Iowa State University, and Peking
University  }}  \maketitle
\begin{abstract}
We consider testing regression coefficients in high dimensional
generalized linear models. An investigation of the test of
\cite{Goeman_Houwelingen_Finos_2011_Biom}  is conducted, which reveals that if the
 inverse of the link
function is unbounded,
the high dimensionality in the covariates can impose  adverse impacts
on the power of the test. We propose a
test formation which can avoid the
 adverse impact of the high dimensionality.
When the inverse of the link function is bounded such as the
logistic or probit regression, the proposed test is 
as good
as \cite{Goeman_Houwelingen_Finos_2011_Biom}'s test.
The proposed tests provide p-values for testing significance for gene-sets as demonstrated in
a case study on  an acute lymphoblastic leukemia dataset.

\vspace*{.3in} \noindent{\bf{Key words}}: Generalized Linear Model;
Gene-Sets; High Dimensional Covariates;   Nuisance Parameter;
$U$-statistics.
\end{abstract}
\end{Large}
\setcounter{section}{1}
 \centerline{\large 1. INTRODUCTION}

\bigskip

\noindent

\def\glm{generalized linear model}
\def\hd{high dimension}

\noindent The generalized linear models
\citep{McCullagh_Nelder_(1989)} are
 widely used statistical models in many fields of statistical applications.
The surge of high dimensional data collection and analysis in
 bioinformatics and related studies have led to the use of  generalized linear models in high dimensional settings. The high dimensionality can
arise at least in two forms. One is in 
the  various multiple response variables but with low or fixed
dimensional covariates where the responses represent the readings
for large number of genes and the covariates represent certain
design and demographic variables. Another is to have  low
dimensional response (for instance  indicators for a disease) but
high dimensional covariates representing genes expressions levels.
Research  works on the first form of high dimensionality
 include \cite{Auer_Doerge_2010}
and \cite{{Lund_2012}} in the context of next generation sequencing data.
 The current paper will be focused on the latter case where the \hd ality is associated with the covariates.
Statistical inference for the \glm s  under the \hd al setting has
been the focus of some latest research.  \cite{Geer_2006_AOS}
considered variable selection  via a LASSO approach.
\cite{Fan_Song_2010_AOS} and \cite{Chang_AOS} proposed
approaches via the sure independence screening of
\cite{Fan_Lv_2008_JRSSB}.

The focus of the paper is on testing for significance of the
regression coefficients of  \hd al generalized linear models, which
is of important interest to practitioners, for instance in the
context of discovering significant gene-sets which is subject to
both \hd ality and multiplicity as the genes in different gene-sets
can overlap. For fixed dimensional data,
 the likelihood ratio test and the Wald
test have been popular choices as elaborated in
\cite{McCullagh_Nelder_(1989)}. {However, the high dimensionality
renders the applicability of these two tests. } There are published
works on testing for the coefficients of high dimensional linear
regression for the large $p$, small $n$ paradigm,
 which include the tests proposed in \cite{Zhong_Chen_2011_JASA} that adapt to the \hd ality and the factorial designs, and in \cite{Lan_Wang_Tsai_2013} that allows testing  on  subsets of the
regression coefficient vector.
\cite{Arias_Candes_Plan_2011_AOS} and
\cite{Ingster_Tsybakov_Verzelen_2010_EJS} studied the higher
criticism tests \citep{Donoho_Jin_2004}
 for  sparse linear regression models, and  demonstrated that the tests can attain the optimal detection boundary for the testing problem.

In an important development,
\cite{Goeman_Houwelingen_Finos_2011_Biom} proposed tests for the
coefficients of high dimensional generalized linear models in the
presence of nuisance parameters.
 The test procedure was formulated by  numerically
simulating a ratio of quadratic forms of certain normally
distributed ``data" to obtain the  critical value.  The test of
\cite{Goeman_Houwelingen_Finos_2011_Biom} allowed the dimension of
the covariates $p$  to be   larger than the sample size $n$,
although $p$ was regarded as fixed. The test has provided a much needed tool
for performing multivariate tests for \glm s where the conventional
likelihood ratio and the Wald tests are not applicable.

In this paper, we first analyze the power properties of
\cite{Goeman_Houwelingen_Finos_2011_Biom}'s test by allowing $p$
diverges to infinity as $n$ increases. It is  discovered that when
the inverse of the link function in the \glm\ is unbounded, for
instance the log link in the Poisson or Negative Binomial
regression,
  the \hd ality can adversely impact the power of the test.   To alleviate the problem,  we propose  tests based on simpler $U$-statistic formulation. 
     A global test for the entire regression coefficient vector and a test for part of the regression coefficients in the presence of  nuisance parameters are proposed.
It is shown that when the inverse of the link function  is bounded,
for instance,  the logistic or the probit link, the proposed test is
equivalent to the test of \cite{Goeman_Houwelingen_Finos_2011_Biom}
asymptotically. However, when the inverse of the link function is
unbounded, as the case of the log link, the proposed tests  have much
better power.  These findings are demonstrated by both theoretical
analysis and numerical simulations.   We apply the proposed tests in
finding significant gene-sets in  an acute lymphoblastic leukemia
dataset.  It is shown in the case study that the p-values produced
from the proposed tests when used in conjunction with a proper
control on the False Discovery Rate
\citep{Benjamini_Hochberg_1995_JRSSB} can lead to finding 
significant gene-sets in the context of \hd ality and multiplicity.

The paper is organized as follows. In Section 2, we review the
inferential setting for the generalized linear models. Section 3
analyzes \cite{Goeman_Houwelingen_Finos_2011_Biom}'s test, which
motivates our proposal for the global test in Section 4 and the test
with nuisance parameters in Section 5.
Results from  simulation studies are reported in Section 6. Section 7 presents the case
study on the acute lymphoblastic leukemia  dataset. All technical details are relegated to the Appendix.

\def\glim{generalized linear model}

 \vskip 1cm

\centerline{\large 2. MODELS AND EXISTING TEST}

\bigskip
\setcounter{equation}{0} \setcounter{section}{2}

\noindent Let $Y$ be a response  variable to a $p$-dimensional
covariate $X$.
{The generalized linear models \citep{McCullagh_Nelder_(1989)}
provide a rich collection of specifications for the conditional mean
of  $Y$ given $X$. Although they are intimately connected to the
exponential family of distributions, a more general  view can be
attained  via the semiparametric quasi-likelihood of Wedderburn
(1974).

Conditioning on the covariate $X$, there exists a monotone function
$g(\cdot)$ and a non-negative function $V(\cdot)$  such that \be
\label{meanvariance} E(Y|X)= \mu(\beta) = g(X^{\T}\beta)\quad
\text{and}\quad \text{var}(Y|X)=V\{g(X^{\T}\beta)\}, \ee where
$\beta$ is a $p$-dimensional coefficient vector and $g^{-1}(\cdot)$
is called the link function.

Let $(X_1,Y_1),\ldots,(X_n,Y_n)$ be the independent copies of
$(X,Y)$ with the first two conditional moments satisfying
(\ref{meanvariance}). The quasi-likelihood of $\beta$ is
\be L_n(\beta) = \sum_{i=1}^n \int_{Y_i}^{\mu_i(\beta
)}\frac{Y_i-t}{V( t )}dt, \label{quasilikelihood} \ee
 where $\mu_i(\beta)=g(X_i^{\T}\beta)$. The maximum quasi-likelihood estimator $\hat{\beta}_n$ of $\beta$
can be obtained by
solving the quasi-likelihood score equation:
\begin{eqnarray}
\ell_n(\beta)=\sum_{i=1}^{n}\frac{ \{ Y_{i}-g (X_{i}^{\T}\beta )\}g'
( X_{i}^{\T}\beta  )X_{i}}{V \{ g( X_{i}^{\T}\beta  )\}}=0.
\label{eq:qscore1}
\end{eqnarray}
The consistency and asymptotic normality of $\hat{\beta}_n$ are well established for fixed dimensional covariate  \citep{McCullagh_Nelder_(1989)}.

Let $\beta = (\beta^{\s1\T}, \beta^{\ss2\T})^{\T}$ be a partition of
the coefficient vector and $X_i=(X_i^{\s1\T}, X_i^{\ss2\T})^{\T}$ be
the corresponding partition of the covariates, where $\beta^{\s1}$
and $X_i^{\s1}$ are $p_1$-dimensional, $\beta^{\ss2}$  and
$X_i^{\ss2}$ are $p_2$-dimensional, and $p_1 + p_2 =p$.
Suppose one is interested in testing a hypothesis
 \be H_{0}:
 \beta^{\ss2}=\beta_0^{\ss2}\quad \text{versus}\quad  H_{1}:  \beta^{\ss2} \neq \beta_0^{\ss2}
 \label{H00}
 \nn
  \ee
on the effect of the second segment of the covariate $X_i^{\ss2}$
while treating $\beta^{\s1}$ as the nuisance parameter.

When the dimensions $p_1$ and $p_2$ are fixed, modified Wald and the
score tests based on the asymptotic Chi-square approximations
\citep{FahrmeirTutz_1994} can be performed to test the above
hypothesis.  However, the latest genomic research
 often requires that $p_2 > n$, see
\cite{Pan_2009}.  When $p_2 > n$,
the conventional Wald or the likelihood ratio tests are no longer
applicable since the invertibility of the information matrix is not
attainable and the maximum likelihood estimators for the parameters
may not be obtained.

\cite{Goeman_Houwelingen_Finos_2011_Biom} considered the following
test formulation in the case of $p_2 > n$ for $g^{-1}(\cdot)$ being
a canonical link. { To make the discussion more generally
applicable, non-canonical links are considered via
$\psi(X_i,\beta_0)=g'(X_i^\T\beta_0)/V\{g(X_i^\T\beta_0)\}$ where
$g'(\cdot)$ and $V(\cdot)$ are, respectively, the first derivative
of  $g(x)$ with respect to $x$ and the
variance function defined in (\ref{meanvariance}).  
The canonical link means  $\psi(X_i,\beta_0) = 1$. Using the general
$\psi(\cdot)$ function does not alter the basic results of
\cite{Goeman_Houwelingen_Finos_2011_Biom}'s test. }

Let $\hat{\beta}_0^{\s1}$ be the estimator of the nuisance parameter
$\beta^{\s1}$ under the null hypothesis,
$\hat{\beta}_0=(\hat{\beta}^{\s1\T}_0,\beta_0^{\ss2\T})^{\T}$,
$\hat{\mu}_{0i} = \mu_i(\hat{\beta}_0)$,
$\widehat{\mu}_0=(\hat{\mu}_{01}, \ldots , \hat{\mu}_{0n})^\T$ and
$\widehat{\bPsi}_0=\{\psi(X_1,\hat{\beta}_0),\ldots,
\psi(X_n,\hat{\beta}_0)\}^{\T}$. Furthermore, let $\bx^{\ss2}=(X_{1}^{\ss2}, \dots,
X_{n}^{\ss2})^{\T}$, $\by=(Y_{1}, \dots, Y_{n})^{\T}$ and $\bd$ is a
$n \times n$ diagonal matrix that collects the diagonal elements of
$\bx^{\ss2}\bx^{\ss2\T}$. The test statistic used in
\cite{Goeman_Houwelingen_Finos_2011_Biom} is \be \label{Goemantest}
\widehat{S}_n=\frac{\{(\by-\widehat{\bmu}_0)\circ
\widehat{\bPsi}_0\}^{\T}
\bx^{\ss2}\bx^{\ss2\T}\{(\by-\widehat{\mu}_0)\circ
\widehat{\bPsi}_0\}}{\{(\by-\widehat{\mu}_0)\circ
\widehat{\bPsi}_0\}^{\T}\bd \{(\by-\widehat{\mu}_0)\circ
\widehat{\bPsi}_0\}}, \ee where the Hadamard product is defined as
$A\circ B=(a_{ij}b_{ij})$ for matrices $A=(a_{ij})$ and
$B=(b_{ij})$.

\setcounter{section}{3} \setcounter{equation}{0}
 \vskip 1cm

\centerline{\large 3. PROPERTIES OF GOEMAN ET AL. (2011)'S TEST }

\bigskip
\noindent We analyze in this section the properties of
the test of \cite{Goeman_Houwelingen_Finos_2011_Biom}.  To make
the discussion focused while being relevance, we concentrate on
testing the global hypothesis
 \[ H_{0}:
 \beta=\beta_0\quad \text{versus}\quad H_{1}\quad  \beta \neq \beta_0
 \]
by assuming $p_2=p$.

To simplify our analysis,  
 we assume $E(X)=0$ without loss of generality as otherwise $X$ can
be re-centered by its mean. Throughout the paper, we denote
$\Sigma_{\sx}=\text{cov}(X)$, $\epsilon=Y-g(X^{\T}\beta)$,
$\epsilon_0=Y-g(X^{\T}\beta_0)$. We use $\|\cdot\|$ to denote the
Euclidean norm, and  for  two sequences $\{a_n\}$ and $\{b_n\}$,
$a_n \asymp b_n$ means $a_n=O(b_n)$ and $b_n=O(a_n)$.

The following assumptions are needed in our analysis.

\begin{as}
There exists a $m$-variate random vector $Z_{i}=(z_{i1}, \dots, z_{i
m})^\T$ for some  $m\ge p$ so that  $ X_{i}=\Gamma Z_{i}$,
 where
$\Gamma$ is a $p\times m$ constant matrix such that $\Gamma
\Gamma^{\T} = \Sigma_{\sx}$ and $E(Z_{i})=0$,
$\text{var}(Z_{i})=\bi_{m}$, where $\bi_{m}$ is the $m \times m$
identity matrix. Each $z_{i j}$ has a finite $8$th moment and
$E(z_{ij}^4)=3+\Delta$ for a constant $\Delta>-3$, and for any
integers $\ell_\nu\ge 0$ and distinct $j_1$, \dots, $j_q$ with
$\sum_{\nu=1}^q\ell_\nu=8$, \be
E(z_{ij_{1}}^{\ell_1}z_{ij_{2}}^{\ell_2}\cdots
z_{ij_{q}}^{\ell_q})=E(z_{ij_{1}}^{\ell_1})E(z_{ij_{2}}^{\ell_2})
\cdots E(z_{ij_{q}}^{\ell_q}).\nn \ee \label{as1}
\end{as}
\begin{as}
\label{as2} As $n \to \infty$, $p \to \infty$,
$\text{tr}(\Sigma_\sx^2)\to \infty$ and
$\text{tr}(\Sigma_\sx^4)=o\{\text{tr}^2(\Sigma_\sx^2)\}$.
\end{as}
\begin{as}
\label{as3} Let $f_{x}$ be the probability density of $X$ and
$D(f_{x})$ be its support. There exist positive constants $K_{1}$
and $K_{2}$ such that $E(\epsilon^{2}|X=x)>K_{1}$ and
$E(\epsilon^{8}|X=x)<K_{2}$ for any $x \in D(f_{x})$.
\end{as}

\begin{as}
\label{as4} $g(\cdot)$ is  once continuous differentiable,
$V(\cdot)>0$, and there exist positive constants $c_{1}$ and $c_{2}$
such that $c_{1}\leq \psi^2(x,
\beta_0)=g'^2(x^{\T}\beta_0)/V^2\{g(x^{\T}\beta_0)\}\leq c_{2}$ for
any $x \in D(f_{x}).$
\end{as}
Assumption \ref{as1} is used in \cite{Bai_Sara_1996_Sinica}
and \cite{Zhong_Chen_2011_JASA} to facilitate
the analysis in ultra high dimensional tests for the means and
linear regression. The model contains the Gaussian and some other
important multivariate distributions as special cases;  see \cite{Chen_Peng_Qin_2009_Biom}.
Assumption \ref{as2} is a weaker substitute to conditions which are
explicit on the relative  rates between  $p$ and $n$, for instance,
$\log(p) \asymp n^{1/3}$, say.
It is noted that when all the eigenvalues of $\Sigma_\sx$ are
bounded, $\text{tr}(\Sigma_\sx^4)=o\{\text{tr}^2(\Sigma_\sx^2)\}$ is
true for any diverging $p$. The condition allows diverging
eigenvalues.
Assumption \ref{as3} is  standard in the analysis of generalized
linear models, for instance, the assumption G in
\cite{Fan_Song_2010_AOS}. In particular, Assumption \ref{as4} is
 satisfied if $Y$ is from the exponential family with canonical
 links. 

For the global hypothesis case,  ${\bmu}_0 =({\mu}_{0 1}, \dots,
{\mu}_{0 n})^{\T}$,  $\bPsi_0=\{\psi(X_1,\beta_0),\ldots,
\psi(X_n,\beta_0)\}^{\T}$ and ${\mu}_{0 i} =g(X_i^{\T}\beta_0)$. The statistic $\widehat{S}_n$ can be written as 
\be \label{Goemantest1} \widehat{S}_n  =  1 + {U_n}/{A_n}, \ee where
$$A_n =
\frac{1}{n}\sum_{i=1}^n\{(Y_i - \mu_{0i})^2\psi^2(X_i,\beta_0)
X_i^{\T} X_i\}\quad \text{and}$$
\[ U_n=
\frac{1}{n}\sum\limits_{i\neq
j}^{n}\{(Y_i-\mu_{0i})(Y_j-\mu_{0j})\psi(X_i,\beta_0)\psi(X_j,\beta_0)X_i^\T
X_j\}.
\]
To facilitate the analysis, we define three matrices: 
$$
\Delta_{\beta,\beta_{0}}= E[\{ g( X^{\T}\beta)-g( X^{\T}\beta
_{0})\}\psi(X,\beta_0)X],$$
 $$\Sigma_{\beta}(\beta_0)
=E[V\{g(X^{\T}\beta)\}\psi^2(X,\beta_0)XX^{\T}]\quad\text{and}
$$
$$
\Xi_{\beta,\beta_0}=E[\{g( X^{\T}\beta)-g(X^{\T}\beta
_{0})\}^2\psi^2(X,\beta_0)XX^{\T}].
$$
For the \glm s,  the difference between $\beta$ and $\beta_0$ is
only detectable through that between $g(X^{\T}\beta)$ and
$g(X^{\T}\beta_0)$. The latter is reflected by
$\Delta_{\beta,\beta_{0}}$  and $\Xi_{\beta,\beta_0}$ defined above.

 Let $\mu_{\sa}$ and
$\mu_{\su}$ be the expectations, $\sigma_{\sa}^2$ and
$\sigma_{\su}^2$ be the variances of $A_n$ and $U_n$ respectively.
Lemma \ref{mean_variance_gt} in the Appendix 
shows that \bea \label{mean}
\mu_{\sa} =\text{tr}\{\Sigma_{\beta}(\beta_0)+\Xi_{\beta,\beta_0}\},  
\quad
\mu_{\su}=(n-1)\Delta_{\beta,\beta_{0}}^{\T}\Delta_{\beta,\beta_{0}},\eea
 \[ \sigma_{\sa}^2 =
{n^{-1}}\left[E\{\epsilon_0^4\psi^4(X,\beta_0)(X^\T
X)^2\}-E^2\{\epsilon_0^2\psi^2(X,\beta_0)(X^\T X)\}\right]\quad
\text{and}\] \be
\sigma_{\su}^2={4(n-2)(1-n^{-1})}{}\xi_{1}+{2(1-n^{-1})}\xi_{2}
\label{variaceU}\ee where $\xi_{1} =
\Delta_{\beta,\beta_{0}}^{\T}\{\Sigma_{\beta}(\beta_{0})+
\Xi_{\beta,\beta_{0}}\}\Delta_{\beta,\beta_{0}} - (\Delta_{\beta,\beta_{0}}^{\T}\Delta_{\beta,\beta_{0}})^{2}  
$~and~  $\xi_{2} =
\text{tr}\{\Sigma_{\beta}(\beta_{0})+\Xi_{\beta,\beta_{0}}\}^2-(\Delta_{\beta,\beta_{0}}^{\T}\Delta_{\beta,\beta_{0}})^{2}.$

From the central limit theorem,
\[
\sigma_\sa^{-1}(A_n-\mu_\sa)\to N(0,1)
\]
in distribution as $n\to\infty.$
By the Taylor expansion,  
\be
\begin{split}
\widehat{S}_n
&= 1+\mu_\sa^{-1} \left\{\mu_\su+(U_n-\mu_\su)\right\} \left\{1-\mu_\sa^{-1}({A_n-\mu_\sa})+\mu_\sa^{-2}({A_n-\mu_\sa})^2+\cdots \right\}  \\
&=
1+\mu_\sa^{-1}\mu_\sb-\mu_\sa^{-2}\mu_\sb(A_n-\mu_\sa)+\mu_\sa^{-1}(U_n-\mu_\su)+\mu_\sa^{-3}\mu_\sb(A_n-\mu_\sa)^2+\cdots.
\end{split}
\label{expansion} \ee

To identify the leading order term  of the above expansion, we consider two families of alternative $H_1$. One is the so-called ``local" alternatives: 
\be \begin{split} \mathscr{L}_{\beta}=\bigg\{\beta_0 \in R^p
~&\bigg|~
\Delta_{\beta,\beta_{0}}^{\T}\Sigma_\sx\Delta_{\beta,\beta_{0}}=o\{n^{-1}\text{tr}(\Sigma^2_\sx)\}~~\text{and}\\
&~~~~~~~~~~~~~~~~~~~~~~~~
\{g(X^{\T}\beta)-g(X^{\T}\beta_{0})\}^{2}=O(1) ~~\text{almost
surely} \bigg\}; \label{eq:local1} \end{split}\ee and the other is
the so-called ``fixed" alternatives: \be
\mathscr{L}_{\beta}^{\mathrm{F}}=\left\{\beta_0 \in
R^p~\bigg|~\Delta_{\beta,\beta_0}^{\T}\Xi_{\beta,\beta_0}\Delta_{\beta,\beta_0}=o\{n^{-1}\text{tr}(\Xi_{\beta,\beta_0}^2)
\}~~\text{and}~~\text{tr}(\Sigma_{\sx}^2)=o\{\text{tr}(\Xi_{\beta,\beta_0}^2)\}
\right\}. \label{globalfix} \ee

It is noted that the null hypothesis $H_0$ is embedded in the
``local" alternatives $\mathscr{L}_{\beta}$. While
$\mathscr{L}_{\beta}$ largely encompasses $\beta$   where the
difference $\|\Delta_{\beta,\beta_0}\|$ is relatively small, it also
includes  a range of  $\beta_0$  not necessarily close to
$\beta$ when  $g(\cdot)$ is uniformly bounded so that
$g(X^{\T}\beta)-g(X^{\T}\beta_{0})$ is uniformly bounded; for
instance,  the logistic or the probit link. 
 It
is a little misleading to call the latter situation as  the ``local"
alternatives  since $\beta_0$ can be quite away from $\beta$ for the
case of $g(\cdot)$  being uniformly  bounded.  We use the term
``local" simply because $H_0$ is part of $\mathscr{L}_{\beta}$.

\def\asy{asymptotic}

 Let $\lambda_1 \le \lambda_2 \cdots \le \lambda_p$ be
 the eigenvalues of $\Sigma_\sx$  and $\lambda_{m_0}$ be the smallest non-zero one for a $m_0 \in \{ 1, \ldots, p\}$.  Since $\Delta_{\beta,\beta_{0}}^{\T}\Sigma_\sx\Delta_{\beta,\beta_{0}} \le \lambda_p \| \Delta_{\beta,\beta_{0}} \|^2$ and $ \text{tr}(\Sigma^2_\sx) \ge \lambda^2_{m_0} (p-m_0)$,   a sufficient condition that ensures the first component of $\mathscr{L}_{\beta}$ is
{ \be \| \Delta_{\beta,\beta_{0}} \|^2 =o \{ \lambda_p^{-1}
\lambda_{m_0}^{2} n^{-1} (p- m_0) \}. \label{localsufficient} \ee

The implication of the ``fixed" alternatives $\mathscr{L}_{\beta}^{\mathrm{F}}$ can be found similarly. Specifically, 
{ if $\{g(X^{\T}\beta)-g(X^{\T}\beta_0)\}^2\asymp \eta(p, n)$ almost
surely for a diverging sequence $\eta(p, n)\to \infty$, the second
component of $\mathscr{L}_{\beta}^{\mathrm{F}}$ is satisfied and a
sufficient condition to the first component of
$\mathscr{L}_{\beta}^{\text{F}}$ is
\[
\| \Delta_{\beta,\beta_{0}} \|^2 =o \{ \lambda_p^{-1}
\lambda_{m_0}^{2} n^{-1} (p- m_0)\eta(p, n) \},
\]
which prescribes a larger magnitude of $\| \Delta_{\beta,\beta_0}
\|^2$ than that 
 in (\ref{localsufficient}) under the ``local"
$\mathscr{L}_{\beta}$. It is noticed that
$\mathscr{L}_{\beta}^{\mathrm{F}}$ is applicable to 
models with unbounded $g(\cdot)$ function such as Poisson or
Negative Binomial regression.

If $\beta_0 \in  \mathscr{L}_{\beta}$,
the proof of Theorem 1 shows that
\[
\mu_\sa=\text{tr}\{\Sigma_{\beta}(\beta_0)+\Xi_{\beta,\beta_0}\},\quad
\mu_{\su}=(n-1)\Delta_{\beta,\beta_{0}}^{\T}\Delta_{\beta,\beta_{0}},
\]
\[
\sigma^2_{\sa}=O(n^{-1}\mu_{\sa}^2),\quad
\sigma^2_{\su}=2\text{tr}\{\Sigma_{\beta}(\beta_0)+\Xi_{\beta,\beta_0}\}^2\{1+o(1)\},
\]
\be
\widehat{S}_n=
1+\mu_\sa^{-1}\mu_\sb+\mu_\sa^{-1}(U_n-\mu_\su)+o_p(\mu_{\sa}^{-1}\sigma_{\su}),
\label{expansion1} \ee and the leading order 
variance of
$\widehat{S}_n$ is
$$\sigma^2_{\widehat{S}_n}=2\text{tr}\{\Sigma_{\beta}(\beta_0)+\Xi_{\beta,\beta_0}\}^2\text{tr}^{-2}\{\Sigma_{\beta}(\beta_0)+\Xi_{\beta,\beta_0}\}.$$
 The following theorem establishes  the asymptotic normality of $\widehat{S}_n$.

\begin{tm}
\label{tmGoeman_local} Suppose Assumptions \ref{as1}-\ref{as4} hold,
then under the ``local" alternatives $ \mathscr{L}_{\beta}$,
\[
\sigma^{-1}_{\widehat{S}_n}(\widehat{S}_n-1-\mu_{\sa}^{-1}\mu_{\su})\to
N(0,1)
\]
in distribution as $n\to\infty.$
\end{tm}
Under the null hypothesis, $\mu_{\sa}^{-1}\mu_{\su} =0$ and
$\sigma^2_{\widehat{S}_n} =
2\text{tr}\{\Sigma^2_{\beta_0}(\beta_0)\}\text{tr}^{-2}\{\Sigma_{\beta_0}(\beta_0)\}$.
Let \[
\widehat{\text{tr}\{\Sigma_{\beta_0}(\beta_0)\}}=\frac{1}{n}\sum_{i=1}^{n}\bigl[\{Y_i-g(X_i^{\T}\beta_0)\}^2\psi^2(X_i,\beta_0)(X_{i}^{\T}X_{i})\bigr]
\quad  \text{and} \]
 \bea
 \label{varestimation}
\widehat{\text{tr}\{\Sigma^2_{\beta_0}(\beta_0)\}}=\frac{1}{n(n-1)}\sum\limits_{i
\neq
j}^{n}\bigl[\{Y_i-g(X_i^{\T}\beta_0)\}^2\{Y_j-g(X_j^{\T}\beta_0)\}^2\psi^2(X_i,\beta_0)\psi^2(X_j,\beta_0)(X_{i}^{\T}X_{j})^{2}\bigr]\eea
be estimators of  $\text{tr}\{\Sigma_{\beta_0}(\beta_0)\}$ and
$\text{tr}\{\Sigma^2_{\beta_0}(\beta_0)\}$, respectively.  Lemma
\ref{pn1} in the Appendix shows that both estimators  are ratioly
consistent under $H_0$.

 Theorem 1 implies an \asy
~$\alpha$-level test that
 rejects $H_{0}$  if
\be
\widehat{S}_n>1+z_{\alpha}\bigl[2\widehat{\text{tr}\{\Sigma^2_{\beta_0}(\beta_0)\}}/\widehat{\text{tr}^2\{\Sigma_{\beta_0}(\beta_0)\}}\bigr]^{1/2},
\label{eq:geomantest} \ee where $z_{\alpha}$ is the upper
$\alpha$-quantile of $N(0,1)$.

\cite{Goeman_Houwelingen_Finos_2011_Biom}  approximated
the null distribution of $\widehat{S}_n$ by simulating repeatedly
versions of $\widehat{S}_n$  by generating $\by$  from a multinormal
distribution $N(\widehat{\bmu}_0,\widehat{\Sigma}_0)$,
 where $\widehat{\Sigma}_0$ is a diagonal matrix
with the $i$th diagonal element being $V(\hat{\bmu}_{0i})$.
A R package ``globaltest" is available at www.bioconductor.org to
implement the algorithm.
The numerical procedure of 
\cite{Goeman_Houwelingen_Finos_2011_Biom} for finding the critical
value is  asymptotically equivalent test procedure to
 that given in (\ref{eq:geomantest}). 

Define the power of the test in (\ref{eq:geomantest}) under the
``local" alternatives $\mathscr{L}_{\beta}$  as \be \Omega_{\sg}
(\beta,\beta_0)=
\text{pr}\left(\widehat{S}_n>1+z_{\alpha}\bigl[2\widehat{\text{tr}\{\Sigma^2_{\beta_0}(\beta_0)\}}/\widehat{\text{tr}^2\{\Sigma_{\beta_0}(\beta_0)\}}\bigr]^{1/2}~\big|~
\beta_0 \in \mathscr{L}_{\beta}\right). \nn \ee The leading order
 power is depicted in the following corollary to  the
\asy\ normality given in Theorem 1 and Lemma \ref{pn1} in the
Appendix.

\begin{cy}
\label{col1} Suppose Assumptions \ref{as1}-\ref{as4} hold, then
\be \Omega_{\sg}( \beta, \beta _{0}) = \Phi \left(-z_{\alpha
}+\frac{n\|\Delta_{\beta,\beta_{0}}\|^2}{\bigl[2\mathrm{tr}\{\Sigma_{\beta}(\beta_0)+\Xi_{\beta,\beta_0}\}^2\bigr]^{1/2}}
\right)\{1+o(1)\} \quad\mbox{ as $n \to \infty$.}
 \nn \ee
\end{cy}

The corollary shows  that the power of
\cite{Goeman_Houwelingen_Finos_2011_Biom}'s test  is determined by
\be
 \text{SNR}(\beta, \beta_0)=\frac{n\|\Delta_{\beta,\beta_{0}}\|^2}{\bigl[2\mbox{tr}\{\Sigma_{\beta}(\beta_0)+\Xi_{\beta,\beta_0}\}^2\bigr]^{1/2}}.\nn
\ee We note that  $\|\Delta_{\beta,\beta_{0}}\|^2$  measures  the
difference
 between $H_{0}$ and $H_{1}$,
 and can be viewed as the signal of the test problem.
At the same time,
$\bigl[2\text{tr}\{\Sigma_{\beta}(\beta_0)+\Xi_{\beta,\beta_0}\}^2\bigr]^{1/2}$
can be regarded as  the noise due to its close connection to the
standard deviation of $\widehat{S}_n$.

Let $\tilde{\lambda}_1\le
\tilde{\lambda}_2\le\cdots\le\tilde{\lambda}_p$ be the eigenvalues
of $\Sigma_{\beta}(\beta_0)+\Xi_{\beta,\beta_0}$. Assumption
\ref{as3} and $\mathscr{L}_{\beta}$ imply  that each
$\tilde{\lambda}_i$ is bounded below and above by constant
multiplies of $\lambda_i$.  Using the same argument leading to
(\ref{localsufficient}),  we can show that $\text{SNR}(\beta,
\beta_0)$ is bounded
 within
$$ \bigl( n \|\Delta_{\beta,\beta_{0}}\|^2 \{2 \tilde{\lambda}_{p}^2 (p-m_0)\}^{-1/2},~~ n \|\Delta_{\beta,\beta_{0}}\|^2\{2 \tilde{\lambda}_{m_0}^2 (p-m_0)\}^{-1/2}\bigr).$$
Thus, if $\|\Delta_{\beta,\beta_{0}}\|$ is a larger order than
$n^{-1/2} {\lambda}_p^{1/2} (p-m_0)^{1/4}$,  $\text{SNR}(\beta,
\beta_0) \to + \infty$ and hence the power converges to 1. { If
$\|\Delta_{\beta,\beta_{0}}\|$ is a smaller order (weaker) than
$n^{-1/2} {\lambda}_{m_0}^{1/2} (p-m_0)^{1/4}$,
the test does not have power beyond the significant level $\alpha$.
Non-trivial power $\Omega_{\sg}(\beta, \beta _{0})$ is attained if
$\|\Delta_{\beta,\beta_{0}}\| \asymp n^{-1/2} {\lambda}_p^{1/2}
(p-m_0)^{1/4}$.

Let  us now evaluate the power of
\cite{Goeman_Houwelingen_Finos_2011_Biom}'s test under the ``fixed"
alternatives $\mathscr{L}_{\beta}^{\mathrm{F}}$, which is denoted as
\[
\Omega_{\sg}^{\mathrm{F}}(\beta,\beta_0)=
\text{pr}\left(\widehat{S}_n>1+z_{\alpha}\bigl[2\widehat{\text{tr}\{\Sigma^2_{\beta_0}(\beta_0)\}}/\widehat{\text{tr}^2\{\Sigma_{\beta_0}(\beta_0)\}}\bigr]^{1/2}~\big|~
\beta_0 \in \mathscr{L}_{\beta}^{\mathrm{F}}\right).
 \]

Unlike the ``local" alternatives case where  $1+ \mu_\sa^{-1}\mu_\sb
+ \mu_\sa^{-1}(U_n-\mu_\su)$ is the leading order term of
$\widehat{S}_n$ in (\ref{expansion}),  the leading order terms under
the ``fixed" alternatives involve an additional term
$\mu_\sa^{-2}\mu_\sb(A_n-\mu_\sa)$.
Note that, having $\mu_\sa^{-2}\mu_\sb(A_n-\mu_\sa)$  does not lead
to  more signal (mean) for the test, but
  can
increase the variance and hence causes a reduction in  the power.
To make this point clear, we consider a specific case where \be
\label{conditionTh3} \|\Delta_{\beta,\beta_{0}}\|^2\asymp
n^{\delta-1}\text{tr}^{1/2}(\Xi^2_{\beta,\beta_0})~~\text{and}~~
\frac{E[\{g(X^{\T}\beta)-g(X^{\T}\beta_0)\}^4\psi^4(X,\beta_0)(X^{\T}X)^2]}{E^2[\{g(X^{\T}\beta)-g(X^{\T}\beta_0)\}^2\psi^2(X,\beta_0)(X^{\T}X)]}\asymp
n^{1-2\delta} \ee for a $\delta \in (0, 1/2)$.  Let
$\tau^2=(\mu_{\su}^2\sigma_{\sa}^2)/(\mu_{\sa}^2\sigma_{\sb}^2)$.

We need one more assumption analogous to Assumption \ref{as2} in the
following analysis.}
 \begin{as}
 \label{as9}
As $n \to \infty$, $p \to \infty$,
$\text{tr}(\Xi_{\beta,\beta_0}^2)\to \infty$ and
$\text{tr}(\Xi_{\beta,\beta_0}^4)=o\{\text{tr}^2(\Xi_{\beta,\beta_0}^2)\}$.
 \end{as}

\begin{tm}
\label{tmGoeman_fixed} Under Assumptions \ref{as1}-\ref{as9}, if
$\{g(X^{\T}\beta)-g(X^{\T}\beta_0)\}^2=O(n^{{1}/{4}})$ almost surely
and (\ref{conditionTh3}) is satisfied, then
 \be
~\Omega_{\sg}^{\mathrm{F}}(\beta,\beta_0)= \Phi
\left(\frac{1}{(1+\tau^2)^{1/2}}\biggl[-z_{\alpha
}+\frac{n\|\Delta_{\beta,\beta_{0}}\|^2}{\{2{\mathrm{tr}}(\Xi^2_{\beta,\beta_0})\}^{1/2}}
\biggr]\right)\{1+o(1)\}\label{eq:power2} \ee as $n\to \infty$, and
$\tau\in (0,\infty)$ is a constant.
\end{tm}

The reason for obtaining the power expression in (\ref{eq:power2})
is that under the conditions of Theorem 2,
$\sigma_{\sa}^2=O(n^{-2\delta}\mu_{\sa}^2)$,
$\sigma_{\su}^2=2\text{tr}(\Xi_{\beta,\beta_0}^2)\{1+o(1)\}$ and \be
\label{fixedone} \widehat{S}_n=
1+\mu_\sa^{-1}\mu_\sb-\mu_\sa^{-2}\mu_\sb(A_n-\mu_\sa)+\mu_\sa^{-1}(U_n-\mu_\su)+o_p(\mu_{\sa}^{-1}\sigma_{\su}).
\ee
Note that, both $\mu_\sa^{-2}\mu_\sb(A_n-\mu_\sa)$ and
$\mu_\sa^{-1}(U_n-\mu_\su)$ are the joint leading order terms of
$\widehat{S}_n$.
The role of Condition (\ref{conditionTh3}) 
 is to make the quadratic terms and beyond  in the Taylor
expansion  (\ref{expansion}) of $\widehat{S}_n$  are of smaller
orders of the two linear terms in (\ref{fixedone}).
A consequence of having $A_n$  in the leading order term  leads to
$\tau^2$ appeared in the power function, which implies a power
reduction. 


 {If the second part of (\ref{conditionTh3}) is more relaxed so that it is of a larger order than $n^{1-2\delta}$ but  a smaller order than $n^{1-\delta}$,  the power expression (\ref{eq:power2}) still holds but with  $\tau^2\to\infty$. This means a dramatic deterioration in the power. 
If the order of the second term in  (\ref{conditionTh3})   is higher
than $n^{1-\delta}$, the quadratic terms and beyond in the expansion
(\ref{expansion}) will be of larger orders than the linear terms in
(\ref{fixedone}), making the power analysis much harder to
accomplish.

\setcounter{section}{4} \setcounter{equation}{0}
 \vskip 1cm

\centerline{\large 4. A NEW PROPOSAL}

\bigskip

\noindent An important insight we have acquired in the analysis of
\cite{Goeman_Houwelingen_Finos_2011_Biom}'s test in the previous
section is that the $A_n$ term in the statistic
$$\widehat{S}_n = 1 + U_n / A_n$$
does not  contribute to the signal of the test but can increase the
variance (noise) and hence adversely affect the power.  Although
$A_n$ has a negligible effect on  the power under the ``local"
alternatives $\mathscr{L}_{\beta}$,  its role on the power becomes
more pronounced under the ``fixed" alternatives
$\mathscr{L}_{\beta}^{\mathrm{F}}$.
 Dividing $A_n$ is a standard formulation that dates back to the Fisher's F-test for regression coefficients.  However, under the high dimensionality, doing so may not be necessary since its contribution to the variance (noise) can be significant as shown in Theorem \ref{tmGoeman_fixed}. 

Our analysis in the previous section
leads us to propose a  statistic by excluding $A_n$ from the
statistic $\widehat{S}_n$. Specifically, we consider a statistic
\[ U_n=
\frac{1}{n}\sum\limits_{i\neq
j}^{n}\{(Y_i-\mu_{0i})(Y_j-\mu_{0j})\psi(X_i,\beta_0)\psi(X_j,\beta_0)X_i^\T
X_j\}.
\]
Comparing with the involved expansion (\ref{expansion}) of
$\widehat{S}_n$, $U_n$ has a much simpler form.  
However, it captures the signal of the test  since $E(U_n) =(n-1) \|
\Delta_{\beta, \beta_0} \|^2$ as  shown in (\ref{mean}). We will
demonstrate in this section that a test based on $U_n$ achieves
better power   than \cite{Goeman_Houwelingen_Finos_2011_Biom}'s test
under $\mathscr{L}_{\beta}^{\mathrm{F}}$ while maintaining the same
\asy\ power under $\mathscr{L}_{\beta}$.

     We  consider testing the global hypothesis $H_0: \beta=\beta_0$
in this section.  A test proposal for the presence of  the  nuisance
parameters  will be unveiled in the next section.

Recall from (\ref{variaceU}),
the variance of $U_n$ is
$$\sigma_{\su}^2={4(n-2)(1-n^{-1})}\xi_{1}+{2(1-n^{-1})}\xi_{2}
$$
where $\xi_{1} =
\Delta_{\beta,\beta_{0}}^{\T}\{\Sigma_{\beta}(\beta_{0})+
\Xi_{\beta,\beta_{0}}\}\Delta_{\beta,\beta_{0}} - (\Delta_{\beta,\beta_{0}}^{\T}\Delta_{\beta,\beta_{0}})^{2}$  
 and 
$\xi_{2} =
\text{tr}\{\Sigma_{\beta}(\beta_{0})+\Xi_{\beta,\beta_{0}}\}^2-(\Delta_{\beta,\beta_{0}}^{\T}\Delta_{\beta,\beta_{0}})^{2}.$
Lemma \ref{variance_local} in the Appendix shows that the \asy\
variance of $U_{n}$ under the ``local" alternatives
$\mathscr{L}_{\beta}$ is
$$
\sigma_{\su}^2=2\text{tr}\{\Sigma_{\beta}(\beta_{0})+\Xi_{\beta,\beta_0}\}^2\{1+o(1)\}.
$$


\begin{tm}
\label{tm1} Suppose Assumptions \ref{as1}-\ref{as4} hold, then under
the ``local" alternatives $ \mathscr{L}_{\beta}$, \be
\frac{U_{n}-n\|\Delta_{\beta,\beta_{0}}\|^2}{\bigl[2\mathrm{tr}\{\Sigma_{\beta}(\beta_{0})+\Xi_{\beta,\beta_0}\}^2\bigr]^{1/2}}
\to N(0,1)  \nn \ee in distribution as $n \to \infty $.
\end{tm}
Theorem \ref{tm1} implies that under the null hypothesis, \be
\frac{U_{n}}{\bigl[2\text{tr}\{\Sigma^2_{\beta_0}(\beta_{0})\}\bigr]^{1/2}}
\to N(0,1)\nn\ee in distribution as $n \to \infty $.
If we use $\widehat{\text{tr}\{\Sigma_{\beta_0}^2 (\beta_0) \}}$ given in  (\ref{varestimation}) to estimate $\text{tr}\{\Sigma_{\beta_0}^2 (\beta_0)\}$, 
 the proposed \asy \
$\alpha$-level test   rejects $H_{0}$  if \be
U_{n}>z_{\alpha}\bigl[2\widehat{\text{tr}\{\Sigma_{\beta_0}^2
(\beta_0) \}}\bigr]^{1/2}. \label{proposetest} \ee

 Let $\Omega
(\beta,\beta_0)$ be the power of the above test  under the ``local"
alternatives $\mathscr{L}_{\beta}$ defined as
\be \Omega (\beta,\beta_0)=
\text{pr}\left(U_{n}>z_{\alpha}\bigl[2\widehat{\text{tr}\{\Sigma_{\beta_0}^2
(\beta_0) \}}\bigr]^{1/2}~\big|~ \beta_0 \in
\mathscr{L}_{\beta}\right).\nn \ee

\begin{cy}
\label{col2} Suppose Assumptions \ref{as1}-\ref{as4} hold, then
\be \Omega( \beta, \beta _{0}) = \Phi \left(-z_{\alpha
}+\frac{n\|\Delta_{\beta,\beta_{0}}\|^2}{\bigl[2\mathrm{tr}\{\Sigma_{\beta}(\beta_{0})+\Xi_{\beta,\beta_0}\}^2\bigr]^{1/2}}
\right)\{1+o(1)\} \quad \mbox{as $n \to \infty $.} \nn \ee
\end{cy}

We note here that the power of the proposed test is \asy ally
equivalent to $\Omega_{\sg}( \beta, \beta _{0})$ of
\cite{Goeman_Houwelingen_Finos_2011_Biom} given in Corollary
\ref{col1}. This is expected since in  the case of ``local"
alternatives $\mathscr{L}_{\beta}$,
$$1+\mu_\sa^{-1}\mu_\sb+\mu_\sa^{-1}(U_n-\mu_\su)$$
  is the leading order term of $\widehat{S}_n$. Hence, the two tests are asymptotically equivalent. 

From Theorem \ref{tm2}, the asymptotic variance of $U_{n}$ under the
``fixed" alternatives  $\mathscr{L}_{\beta}^{\mathrm{F}}$ is
$$
\sigma_{\su}^2=2\text{tr}(\Xi^2_{\beta,\beta_0})\{1+o(1)\}.
$$

Let $\Omega^{\mathrm{F}} (\beta,\beta_0)$ be the power of the
proposed test under the ``fixed" alternatives
$\mathscr{L}_{\beta}^{\mathrm{F}}$ given by

\be \Omega^{\mathrm{F}} (\beta,\beta_0)=
\text{pr}\left(U_{n}>z_{\alpha}\bigl[2\widehat{\text{tr}\{\Sigma_{\beta_0}^2
(\beta_0) \}}\bigr]^{1/2}~\big|~ \beta_0 \in
\mathscr{L}_{\beta}^{\mathrm{F}}\right).\nn \ee

\begin{tm}
\label{tm2} Suppose Assumptions \ref{as1}-\ref{as9} hold, if
$\{g(X^{\T}\beta)-g(X^{\T}\beta_0)\}^2=O(n^{1/4})$ almost surely,
then
\be \Omega^{\mathrm{F}}(\beta, \beta _{0}) = \Phi \left(-z_{\alpha
}+\frac{n\|\Delta_{\beta,\beta_{0}}\|^2}{\{2\mathrm{tr}(\Xi^2_{\beta,\beta_0})\}^{1/2}}
\right)\{1+o(1)\} \quad \mbox{as $n \to \infty $.}
 \nn \ee
\end{tm}

The conditions in Theorem \ref{tm2} are  simpler than those in
Theorem \ref{tmGoeman_fixed},  as Condition (\ref{conditionTh3}) is
not needed.
To compare the two power functions under the ``fixed" alternatives
while assuming the conditions of Theorem 2,
(\ref{conditionTh3}) implies that 
$$\frac{n\|\Delta_{\beta,\beta_{0}}\|^2}{\{2\text{tr}(\Xi^2_{\beta,\beta_0})\}^{1/2}}\asymp n^{\delta} \to \infty.$$
A power gain of the proposed test is evident as
$\Omega^{\mathrm{F}}(\beta, \beta
_{0})>\Omega_{\sg}^{\mathrm{F}}(\beta,\beta_0)$ asymptotically,
since the power function of
\cite{Goeman_Houwelingen_Finos_2011_Biom}'s test given in
(\ref{eq:power2}) has an extra $\tau^2$ in the denominator.

\setcounter{section}{5} \setcounter{equation}{0}

 \newpage

\centerline{\large 5. TEST WITH NUISANCE PARAMETER}

\bigskip

\noindent We consider testing for parts of the regression
coefficient vector $\beta$.  This is motivated by practical needs to
consider the significance for a subset of covariates, say
$X^{\ss2}$,
 in the presence of other covariates, say $X^{\s1}$.   For instance, one may have both gene expression levels and demographic variables collected in a study on the cause of a disease. The  researcher may be  interested only in the effect of the genes on the disease.
 In this case, the coefficients to the demographic variables may be viewed as nuisance parameters.

Without loss of generality, we partition $\beta=(\beta^{\s1\T},
\beta^{\ss2\T})^\T$  where the dimensions of $\beta^{\s1}$ and
$\beta^{\ss2}$ are respectively $p_1$ and $p_2$. 
It is of interest to test \be H_{01}:
 \beta^{\ss2}=\beta_0^{\ss2}\quad\text{versus}\quad H_{1 1}:  \beta^{\ss2} \neq \beta_0^{\ss2}
 \label{H0nuisance}
 \nn
  \ee
in the presence of the nuisance  $\beta^{\s1}$.

 A test statistic along the line of the global test statistic $U_{n}$ in the previous section will be proposed.
To this end, the nuisance parameter has to be estimated first under
$H_{01}$. 
The quasi-likelihood score of $\beta^{\s1}$ is
\[ \ell_1(\beta^{\s1},\beta^{\ss2})=\frac{\partial
L_{n}(\beta)}{\partial \beta^{\s1}}
=\bx^{\s1\T}\{(\by-\bmu)\circ\bPsi\}
\]
where $L_{n}(\beta)$ is defined in (\ref{quasilikelihood}),
$\bx^{\s1}$ is similarly defined as $\bx^{\ss2}$ in Section 2,
$\bmu=\{\mu_{1}(\beta),  \dots, \mu_{n}(\beta)\}^{\T}$ where
$\mu_{i}(\beta)=g(X_i^{\T}\beta)$  and $\bPsi=\{\psi(X_1,\beta),
\dots, \psi(X_n,\beta)\}^{\T}$. The maximum quasi-likelihood
estimator of $\beta^{\s1}$ under $H_{01}$ solves
$$\ell_1(\beta^{\s1},\beta^{\ss2}_0)=0,$$
which is denoted as $\hat{\beta}^{\s1}_0$. Let
$\hat{\beta}_0=(\hat{\beta}^{\s1\T}_0,\beta_0^{\ss2\T})^{\T}$ and
$\hat{\mu}_{0i} = \mu_i(\hat{\beta}_0)$.

We consider a statistic,
 \be \label{eq: testnui}
\widetilde{U}_n=\frac{1}{n}\sum\limits_{i \neq j}^{n}\{
(Y_{i}-\hat{\mu}_{0i})(Y_{j}-\hat{\mu}_{0j})\psi(X_i,\hat{\beta}_0)\psi(X_j,\hat{\beta}_0)X_{i}^{\ss2\T}X_{j}^{\ss2}\}.
\ee

Let $\Sigma_{\sx^{\mathrm{\scriptscriptstyle
(i)}}}=E(X^{\mathrm{\scriptscriptstyle
(i)}}X^{\mathrm{\scriptscriptstyle (i)}\T})$ for $i=1$ and $2$. The
following assumptions are needed in the analysis of this section.
\begin{as}
\label{as5} As $n \to \infty$, $p_2 \to \infty$,
$\text{tr}(\Sigma_{\sx^{\ss2}}^2)\to \infty$ and
$\text{tr}(\Sigma_{\sx^{\ss2}}^4)=O\{n^{-1}\text{tr}^2(\Sigma_{\sx^{\ss2}}^2)\}$.
\end{as}
\begin{as}
\label{as6} As $n \to \infty$, $p_1n^{-1/4} \to 0$ and there exists
a $\beta^{\ast \s1} \in R^{p_1}$ such that
$\|\hat{\beta}^{\s1}_0-\beta^{\ast \s1}\|=O_p(p_1n^{-1/2})$, and in
particular under $H_{01}$, $\beta^{\ast \s1}=\beta^{\s1}$, where
$\beta=(\beta^{\s1\T}, \beta^{\ss2\T})^\T$ is the true parameter.
\end{as}
\begin{as}
\label{as7} There exists a positive constant $\lambda_0$ such that
$0 <\lambda_0\leq \lambda_{\min}(\Sigma_{\sx^{\s1}})\leq
\lambda_{\max}(\Sigma_{\sx^{\s1}})\leq \lambda_0^{-1}<\infty$, where
$\lambda_{\min}(\Sigma_{\sx^{\s1}})$ and
$\lambda_{\max}(\Sigma_{\sx^{\s1}})$ represent the smallest and
largest eigenvalues of the matrix $\Sigma_{\sx^{\s1}}$ respectively.
\end{as}

\begin{as}
\label{as8}
 $g(\cdot)$, $\psi(\cdot)$ are once continuous differentiable functions, $V(\cdot)>0$ and there exist positive constants $c_{1}$ and $c_{2}$  such that for $\beta^{\ast}_0 = (\beta^{\ast \s1\T}, \beta_0^{\ss2\T})^{\T}$ where $\beta^{\ast \s1}$ is defined in Assumption \ref{as6}, $c_{1}\leq \psi^2(x,\beta^{\ast}_0)=
g'^2(x^{\T}\beta^{\ast}_0)/V^2\{g(x^{\T}\beta^{\ast}_0)\} \leq
c_{2}$, $[\partial \psi\{g(t)\}/\partial
g(t)]^2~|_{t=x^{\T}\beta^{\ast}_0}\leq c_{2}$ for any $x \in
D(f_{x})$ and a neighborhood of $x^{\T}\beta^{\ast}_0$.
\end{as}

These assumptions are variations of Assumptions \ref{as2}-\ref{as4}
in Section 2. Specifically, Assumption \ref{as5} is equivalent to
Assumption \ref{as2} in the presence of the nuisance parameter.  The
requirement of the growing rate of $p_1$ being slower than $n^{1/4}$
is to allow accurate estimation of the nuisance parameter in the
context of \hd ality. {Assumption \ref{as6} maintains that under the
alternative hypothesis, the limit of the quasi-likelihood estimator
$\hat{\beta}^{\s1}_0$  may deviate from  the true parameter
$\beta^{\s1}$,  when the discrepancy between $\beta_0^{\ss2}$ and
$\beta^{\ss2}$ is large. That $\hat{\beta}^{\s1}_0$ converges to a
limit is commonly assumed in  hypothesis  testing under the
alternative hypothesis which is related to the notation of the
maximum likelihood estimation under a misspecified model
\citep{White_1982_Eco}.
Assumption \ref{as7} is easier to be satisfied due to
$\Sigma_{\sx^{(1)}}$'s dimension is much more manageable than the
case considered in the previous section. Assumption \ref{as8} is an
updated version of Assumption \ref{as4} to suit the case of nuisance
parameters.

To analyze the power,  we introduce two matrices 
\be \Delta_{\beta,{\beta}_0^{\ast}}^{\ss2} =
E[\{g(X^{\T}\beta)-g(X^{\T}{\beta}_0^{\ast})\}\psi(X,{\beta}_0^{\ast})X^{\ss2}]\quad
\text{and} \nn\ee \be
 \Sigma_{\beta}^{\ss2}({\beta}_0^{\ast}) =
E[V\{g(X^{\T}\beta)\}\psi^2(X,{\beta}_0^{\ast})X^{\ss2}X^{\ss2\T}],
\nn \ee which are counterparts of $\Delta_{\beta,\beta_0}$ and
$\Sigma_{\beta}(\beta_0)$ used in the study of the global test.
There is no need to define a counterpart of $\Xi_{\beta,\beta_0}$
since the second part of the ``local" alternatives
$\mathscr{L}_{{\beta}^{\ss2}}$ defined below makes it unnecessary.

The involvement of the estimated nuisance parameter
$\hat{\beta}^{\s1}_0$  does complicates the power analysis of the
test.  To expedite the study,  our analysis is confined under the
following family of the  ``local" alternatives
\[
\mathscr{L}_{{\beta}^{\ss2}}=\bigg\{~\beta^{\ss2}_0\in
R^{p_2}~\bigg|~
\Delta_{\beta,{\beta}_0^{\ast}}^{\ss2\T}\Sigma_{\sx^{\ss2}}\Delta_{\beta,{\beta}_0^{\ast}}^{\ss2}=o\{n^{-1}\text{tr}(\Sigma_{\sx^{\ss2}}^2)\}~~\text{and}~~
E\{g(X^{\T}\beta)-g(X^{\T}{\beta}_0^{\ast})\}^4=o(n^{-3/2}) \bigg\}.
\]
We note here that the second component of
$\mathscr{L}_{{\beta}^{(2)}}$ is stronger than that in
$\mathscr{L}_{\beta}$  in (\ref{eq:local1}), which simplifies the
analysis in the presence  of the nuisance parameter.

The asymptotic normality of $\widetilde{U}_n$
 is given in the following theorem.
\begin{tm}
\label{tm3} Under Assumptions \ref{as1}, \ref{as3},
\ref{as5}-\ref{as8}, and  the ``local" alternatives
$\mathscr{L}_{{\beta}^{\ss2}}$,
 \be
\frac{\widetilde{U}_n-n\|\Delta_{\beta,{\beta}_0^{\ast}}\|^2}{\bigl[2\mathrm{tr}\{\Sigma^{\ss2}_{\beta}({\beta}_0^{\ast})\}^2\bigr]^{1/2}}
\to N(0,1)
\nn \ee in distribution as $n \to \infty $.
\end{tm}

To formulate a test procedure from the above \asy\ normality,
we use
\be \widehat{R}_n=\frac{1}{n(n-1)}\sum\limits_{i \neq j
}^{n}(Y_{i}-\hat{\mu}_{0i})^{2}(Y_{j}-\hat{\mu}_{0j})^{2}\psi^2(X_i,\hat{\beta}_0)\psi^2(X_j,\hat{\beta}_0)(X_{i}^{\ss2\T}X_{j}^{\ss2})^{2}
\nn \ee to estimate
$\text{tr}\{\Sigma^{\ss2}_{\beta}({\beta}_0^{\ast})\}^2$ under 
 $H_{0 1}$.
The following proposition shows the estimator is ratioly consistent.

\begin{pn}
\label{tm4} Under Assumptions \ref{as1}, \ref{as3},
\ref{as5}-\ref{as8} and $H_{01}$, 
\[
\frac{\widehat{R}_{n}}{\mathrm{tr}\{\Sigma^{\ss2}_{\beta}({\beta}_0^{\ast})\}^2}\to
1
\]
in probability as $n \to \infty $.
\end{pn}

Hence, an asymptotic $\alpha$-level test  rejects $H_{0 1}$ if
$\widetilde{U}_n>z_{\alpha}(2\widehat{R}_{n})^{1/2}$ and
 the power of the test
under the ``local" alternatives $\mathscr{L}_{{\beta}^{\ss2}}$ is
\[
\Omega^{\ss2}(\beta,
{\beta}_0^{\ast})=\text{pr}\left(\widetilde{U}_n>z_{\alpha}(2\widehat{R}_{n})^{1/2}\mid\beta^{\ss2}_0\in\mathscr{L}_{{\beta}^{\ss2}}\right).
\]
\begin{cy}
\label{cy2} Under Assumptions \ref{as1}, \ref{as3},
\ref{as5}-\ref{as8},
\[
\Omega^{\ss2}(\beta, {\beta}_0^{\ast})=\Phi \left(-z_{\alpha
}+\frac{n\|\Delta^{\ss2}_{\beta,{\beta}_0^{\ast}}\|^2}{\bigl[2\mathrm{tr}\{\Sigma^{\ss2}_{\beta}({\beta}^{\ast})\}^2\bigr]^{1/2}}
\right)\{1+o(1)\} \quad \mbox{as $n \to \infty $.}
\]
\end{cy}
The power  $\Omega^{\ss2}(\beta,{\beta}_0^{\ast})$ has a similar
form as  $\Omega(\beta,\beta_0)$ in Corollary \ref{col2}. This is
expected  due to the close connection between the two tests and
their test statistics respectively.
We note that the denominator inside  $\Phi(\cdot)$  only involves
$\Sigma^{\ss2}_{\beta}({\beta}^{\ast})$ due to
the second part of
 $\mathscr{L}_{{\beta}^{\ss2}}$.

We did not study the power  under a version of the ``fixed"
alternatives  similar to the one defined in Section 3, as we would
expect the power performance would be largely similar to the one
depicted in Section 4 for the proposed global test. We  also did not
study the power property of the
\cite{Goeman_Houwelingen_Finos_2011_Biom}'s test with nuisance
parameter as the analysis would be quite involved due to the
division of $A_n$ term  and the estimated nuisance parameter.
However, we would expect similar power properties as revealed in the
previous section would prevail to the nuisance parameter case,
namely the power performance of
\cite{Goeman_Houwelingen_Finos_2011_Biom}'s test would be hampered
when the inverse of the link function is unbounded.
This is indeed confirmed by the simulation studies reported in the
next section.

\setcounter{section}{6} \setcounter{equation}{0}
 \vskip 1cm

\centerline{\large 6. SIMULATION STUDIES}

\bigskip

\noindent We report in this section  results from simulation studies
which were designed to evaluate the performances of the proposed
high dimensional test procedures for the generalized linear models.
Both the global test and the test in the presence of
nuisance parameter were considered for both the proposed  and
\cite{Goeman_Houwelingen_Finos_2011_Biom}'s tests.

Throughout this section,  
the covariates $X_i=(X_{i 1}, \ldots, X_{i p})^{\T}$ were
 generated according to a moving average model
\be
X_{ij}=\rho_{1}Z_{ij}+\rho_{2}Z_{i(j+1)}+\cdots+\rho_{T}Z_{i(j+T-1)},
\quad  j=1,\ldots,p; \label{eq:s1} \ee for some $T<p$, where
$Z_{i}=(Z_{i1},\ldots,Z_{i(p+T-1)})^{\T}$ were from a $(p+T-1)$
dimensional standard normal distribution $N(0,\bi_{p+T-1})$. The
coefficients $\{\rho_{l}\}_{l=1}^{T}$ 
were generated independently from the $U(0,1)$ distribution, and
were treated as fixed once generated.  Here,  $T$ was used to
prescribe different levels of  dependence among the components of
the \hd al vector $X_{i}$. We had experimented $T=5, 10$ and $20$,
and only reported the results for $T=5$ since those for $T=10$ and
$20$ were largely similar.

Three \glim s were considered in the simulation study: the logistic,
Poisson and Negative Binomial regression models respectively. In the
logistic regression model, the conditional mean of the response $Y$
was given by
\[
 E(Y_i|X_i)= g(X_i^{\T}\beta)=\frac{\text{exp}(X_i^{\T}\beta)}{1+\text{exp}(X_i^{\T}\beta)},
\]
and conditioning on $X_i$, $Y_i \sim
\text{Bernoulli}\{1,g(X_i^{\T}\beta)\}$. In the Poisson regression,
\[
E(Y_i|X_i)=g(X_i^{\T}\beta)=\text{exp}(X_i^{\T}\beta),
\]
and conditioning on $X_{i}$,  $Y_{i} \sim \text{Poisson}\{g(X_i^{\T}
\beta)\}$. The setup for the Negative Binomial model  was
\[
Y|\lambda\sim \text{Poisson}(\lambda)\quad \hbox{and} \quad
\lambda\sim \text{Gamma}\{\text{exp}(X^{\T}\beta), 1\}.
\]
The conditional distribution of $Y$ given $X$ is the negative
binomial distribution $NB\{\text{exp}(X^{\T}\beta), 1/2\}$, which
prescribes an over-dispersion to the  Poisson model, and makes it a
popular alternative to the Poisson regression in practice.

To create regimes of high dimensionality, we chose a relationship
$p=\text{exp}(n^{0.4})$ and specifically considered $(n,p)=(80,320)$
and $(200,4127)$ in the simulations.   Seven nominal type I errors
ranging from 0.05 to 0.2 were considered, and the corresponding
empirical sizes and powers were evaluated from 2000
 replications.

We first considered testing the global hypothesis 
\be \label{eq:sumi1} H_{0}: \beta=0_{p \times
1}\quad\text{versus}\quad H_{1}: \beta \neq  0_{p \times 1}. \ee
 In designing the alternative hypothesis, we made
$\|\beta\|^{2}=2$ and
chose the first five coefficients in $\beta$ to be non-zero of equal
magnitude and the rest of the coefficients to be zero.  Hence, the
non-zero coefficients were quite sparse. In order to have a
reasonable range for the response variable, as in
\cite{Goeman_Houwelingen_Finos_2011_Biom}, we restricted
$E(Y_i|X_i)$  between $\text{exp}(-4)/\{1+\text{exp}(-4)\}=0.02$ and
$\text{exp}(4)/\{1+\text{exp}(4)\}=0.98$ for the logistic model,
and between $\text{exp}(0)=1$ and $\text{exp}(4)=55$ for the Poisson
and Negative Binomial models respectively.

The empirical power profiles (curves of empirical power versus
empirical size)  of the global tests for the three \glim s were
plotted in Figure \ref{Figglobaltest}.
 It is observed that the
proposed global test and \cite{Goeman_Houwelingen_Finos_2011_Biom}'s
test had largely similar power profiles for the logistic model as
displayed by Panels (a) and (b) of the figure.
 This is consistent with
our findings in
 Corollaries \ref{col1} and \ref{tmGoeman_fixed}, which indicate
that both tests have the same asymptotic powers under the ``local"
alternatives $\mathscr{L}_{\beta}$.  It is noted that, the inverse
of the logistic link function
is uniformly bounded and thus
 is covered by
$\mathscr{L}_{\beta}$.  Panels (a) and (b) of Figure
\ref{Figglobaltest} displayed that the proposed test had a slightly
higher power than \cite{Goeman_Houwelingen_Finos_2011_Biom}'s test
in the case of the logistic model. This can be understood as the
impact of $A_n$ term on    the variance of $\widehat{S}_n$ despite
its being the second order only in the case
of the ``local" alternatives.

Panels (c)-(f) of Figure \ref{Figglobaltest} showed  a much larger
discrepancy in the power profiles between the two tests for the
Poisson and Negative Binomial models with the proposed test being
significantly more powerful.
It is noted that both models have  unbounded $g(\cdot)$,
which imply that the testing was operated in the regime of the
``fixed" alternatives $\mathscr{L}_{\beta}^{\text{F}}$. The
simulated power profiles confirmed the findings in Theorem
\ref{tmGoeman_fixed}
in that an unbounded $g(\cdot)$ function can 
adversely impact the power of
\cite{Goeman_Houwelingen_Finos_2011_Biom}'s test, whereas the
proposed test withstands such situations due to its test statistic
formulation.

We then conducted  simulation for testing \be \label{nuisanceH0}
H_{0}: \beta^{\ss2}=0_{p_2 \times 1}\quad \text{versus}\quad H_{1}:
\beta^{\ss2} \neq  0_{p_2 \times 1} \ee in the presence of nuisance
parameter $\beta^{\s1}$ for the same three generalized linear models
considered above. 
The nuisance parameter $\beta^{\s1}$ was  $p_1=10$ dimensional,
generated randomly from $U(0,1)$ as in the design of the global
hypothesis.
 We still
chose $(n,p_2)=(80,320)$ and $(200,4127)$ by assigning
$p_2=\text{exp}(n^{0.4})$. To evaluate the power of the test, the
first five elements of $\beta^{\ss2}$ were set to be non-zero of
equal magnitude with $\|\beta^{\ss2}\|^2=2$ while the rest of
$\beta^{\ss2}$  were zeros.

The power profiles of the proposed and \cite{Goeman_Houwelingen_Finos_2011_Biom}'s tests were displayed in 
Figure \ref{Fignuisance}.
It is observed from Panel (a) of Figure \ref{Fignuisance} that,
 for the logistic model with $n=80$ and $p_2=320$,
 the test of \cite{Goeman_Houwelingen_Finos_2011_Biom} had very
severe size distortion,
 which may be   due to the
estimation of the nuisance parameter.  The test formulation of
\cite{Goeman_Houwelingen_Finos_2011_Biom} required generating $Y\sim
N(\widehat{\mu}_0,\widehat{\Sigma}_0)$ in the approximation of  the
null distribution, which can amplify the estimation errors in the
nuisance parameter case, as compared with the test statistic
$\widetilde{U}_n$.
We observed that when the sample size was increased to $n=200$,
Panel (b) of Figure \ref{Fignuisance} shows the size distortion
 is no longer that
severe as compared with the case of $n=80$. As our test statistic
$\widetilde{U}_n$ does not require regenerating $Y$ based on the
estimated $\widehat{\mu}_0$ and $\widehat{\Sigma}_0$, it is more
robust.  Indeed, the size distortion presence for the test of
\cite{Goeman_Houwelingen_Finos_2011_Biom}
was largely absence for the proposed test.
Figure \ref{Fignuisance} shows that the proposed test had quite
reasonable power with good control of the type I error.  For the
Poisson and Negative Binomial models, we observed that the proposed
test had much more advantageous power profiles than those of
\cite{Goeman_Houwelingen_Finos_2011_Biom}'s test. The latter was
similar to the global tests demonstrated in Figure
\ref{Figglobaltest}.


\setcounter{section}{7} \setcounter{equation}{0}
 \vskip 1cm

\centerline{\large 7.  CASE STUDY}

\bigskip

\noindent We analyze a  dataset that contains  microarray readings
for 128 persons who suffer the acute lymphoblastic leukemia.
 The dataset also has information on   patients'  age,  gender and  response to multidrug resistance.   Among the 128 individuals, 75 of them were patients of the B-cell type leukemia which were classified further to two types: the BCR/ABL fusion $(35~patients)$ and cytogenetically normal NEG  $(40~patients)$.
  The dataset has
been analyzed by \cite{Chiaretti_2004_Blood}, \cite{Dudoit_2008},
\cite{Chen_Qin_2010_AOS} and \cite{Li_Chen_2012_AOS} and others
motivated from different aspects of the inference.

Biological  studies have shown  that each gene tends to work with
other genes to perform
 certain biological missions. Biologists have defined gene-sets under the
Gene Ontology system which provides structured vocabularies
producing names of Gene Ontology terms. The gene-sets under the Gene
Ontology system have been classified to  three broad functional
categories:
 Biological Processes, Cellular
Components and Molecular Functions. There have been a set of
research works focusing on identifying differentially expressed sets
of genes in the analysis of gene expression data; see
\cite{Efron_2007_AOAS}, \cite{Rahmatallah_2012_Bioinformatics}.
After preliminary gene-filtering with the algorithm proposed in
\cite{Gentleman_2005}, there were  2250 unique Gene Ontology terms
in Biological Processes, 328 in Cellular Component and 402 in
Molecular Function categories respectively, which involved 3265
genes in total.


Our aim here is to identify  gene-sets within each functional
category, which are significant in
 determining the two types of B-cell  acute lymphoblastic leukemia:  BCR/ABL fusion or  cytogenetically normal NEG.
We formulate it as a binary regression problem with the response
$Y_i$ being 1 if the $i$th patient  had the  BCR/ABL type
 acute lymphoblastic leukemia and 0 if had the NEG type.  The covariate of the $i$th patient
corresponding to a gene-set, label by $g$ in the subscript, is $X_{i
g} = (X^{\s1\T}_{i g}, X^{\ss2\T}_{i g})^{\T}$, where $X^{\s1}_{ig}$
contains the gender, age and the patient's response to multidrug
resistance (1 if negative and 0 positive),
 and  $X^{\ss2}_{i g}$ is the vector of gene expression levels of the $g$th
Gene Ontology term.

We considered the logistic and probit models
for the gene-set data due to the binary nature of the response
variable.
The two models are, respectively, \be E(Y_i|X^{\s1}_{i
g},X^{\ss2}_{i g})=\frac{\text{exp}(X^{\s1\T}_{i g} \beta^{\s1}_{
g}+X^{\ss2\T}_{i g} \beta^{\ss2}_g)}{1+\text{exp}(X^{\s1\T}_{i
g}\beta^{\s1}_{ g}+X^{\ss2\T}_{i g}\beta^{\ss2}_{ g})}\quad
\text{and}\quad \nn \ee \be E(Y_i|X^{\s1}_{i g},X^{\ss2}_{i
g})=\Phi(X^{\s1\T}_{i g}\beta^{\s1}_g+X^{\ss2\T}_{i g}
\beta^{\ss2}_g). \nn \ee


For the  leukemia data, it is of fundamental interest in discovering
significant Gene Ontology terms while considering the effects of the
three covariates in  $X^{\s1}$, namely by treating $\beta_g^{\s1}$
as the nuisance parameter and testing the following hypothesis: \be
H_0:~\beta^{\ss2}_g=0\quad \text{versus}\quad
H_1:~\beta^{\ss2}_g\neq 0. \label{caseH0}\nn \ee

By controlling the false discovery rate
\citep{Benjamini_Hochberg_1995_JRSSB} at 0.01, 1084 gene-sets in
Biological Processes, 154 in Cellular Components and 153 in
Molecular Function were found significant under the logistic model,
and 981 in Biological Processes, 140 in Cellular Components and 132
in Molecular Function were significant under the probit model.
Table\ref{table2} reports the two by two rejection/non-rejection
classification between the tests under the two models.  It shows
that the testing results were largely agreeable between the two
models. This was especially the case for the gene-set categories of
Biological Processes and Cellular Components, with more than 90\% of
the gene-sets rejected under the logistic model being also rejected
under the probit model, and the non-rejected gene-sets matched
perfectly.  The discrepancy in the test conclusions got larger for
gene-sets in the Molecular Function category. But still, the
percentages of agreement between the two models exceeded 72\% in the
rejection and 92\% in the non-rejection.  These showed again the
testings under the  two models attained similar  results.

We also carried out the global test for the significance of the
entire regression coefficient vector $\beta_g$  by performing test
on
 \be
H_0:~\beta_g=0\quad \text{versus}\quad H_1:~\beta_g\neq 0
\label{caseH0} \nn\ee where
$\beta_g=(\beta_g^{\s1\T},\beta_g^{\ss2\T})^{\T}$ with the first
three coefficients corresponding  to the three non-genetic
covariates: the gender, age and multidrug resistance. We note that
the  value of the standardized global test statistics under the
logistic and the probit models were identical.    This is because
under the $H_0$, $g(X_i^{\T}\beta_g)=g(0)=0.5$  and $\psi(X_i,0)$
are constant for both models,
which means that $\psi(X_{i},0)$ are canceled out in the
standardized test statistics.
 Hence, the
test procedures were identical for testing the global hypothesis
 regarding each gene-set under both the logistic and probit models.

Figure \ref{caseglobal} displays the histograms of p-values and the
standardized global test statistics $L_n$. It is observed that the
bulk of the test statistics (right panels) took extremely large
values in the scale of the standard normal distribution, implying
that most of the p-values would be very small
and the significance of many sets of genes.  The latter was
confirmed by the left panels of  Figure \ref{caseglobal}.  The
histograms of the standardized test statistics and the p-values of
the test for the gene-sets only while treating the first three
coefficients as the nuisance parameter are shown in  Figures \ref{caselogiticnui}
and \ref{caseprobitnui}.
 Comparing Figure \ref{caseglobal} with Figures \ref{caselogiticnui}
and \ref{caseprobitnui},  it is found that the body of the
histograms were much less extreme in Figures \ref{caselogiticnui}
and \ref{caseprobitnui} than those in Figure \ref{caseglobal}.  This
indicates that much of the significance in the global tests were due
to the significance of the three nuisance covariates rather than the
gene-sets.  It also demonstrates that considering the three nuisance
parameters was necessary in filtering out the influence of the
gene-sets
between the two types of B-cell acute lymphoblastic leukemia.

\vskip 1cm

\centerline{\large 8. DISCUSSION}

\bigskip

\noindent As  the \glm s are 
widely used tools 
in analyzing genetic data, the proposed tests,    being more
adaptive to the \hd ality,  are useful additions to the existing
test procedures for the significance of regression coefficients. As
shown in the case study,  testing for the significance of gene-sets
requires \hd al multivariate test procedures which can produce
p-values  under both \hd ality and multiplicity (as genes in
gene-sets can overlap). The proposed tests and the tests of
\cite{Goeman_Houwelingen_Finos_2011_Biom} are such tests which can
be used for the gene-sets testing in conjunction with the FDR
procedure to control the overall family-wise error rate when testing
a large number of hypotheses simultaneously.

The test of \cite{Goeman_Houwelingen_Finos_2011_Biom} was proposed
for fixed dimension $p$ which can be larger than $n$.  The authors
focused much on the size approximation of the test rather than on
the power performance. The proposed tests are designed to  improve
the performance of \cite{Goeman_Houwelingen_Finos_2011_Biom}'s test
in the context of \hd ality.  This is especially the case  when the
inverse of the link function is unbounded, which permits  the \hd
ality to insert adverse influence on the test of
\cite{Goeman_Houwelingen_Finos_2011_Biom}.  The proposed test
statistics due to their simpler formulations can avoid some of the
\hd al effects, and hence lead to better test  performances in terms
of more accurate size approximation and more power in detecting the
significant regression coefficient vector.

\appendix

\setcounter{section}{1} \setcounter{equation}{0}
 \vskip 1cm

\centerline{\large   APPENDIX}

\bigskip

In this section, we provide technical proofs to the main results
reported in Section 3-5. To establish the results of the paper, we
introduce {three}  lemmas whose proofs are available in
\cite{Chen_Guo_2014}. 

\bigskip
\bigskip

\noindent  We define  a few notations:
 \be
\epsilon_i=Y_i-g(X_i^{\T}\beta),\quad
\epsilon_{0i}=Y_i-g(X_i^{\T}\beta_0),\quad
V_{0i}=V\{g(X_i^{\T}\beta_{0})\},\quad
\psi_{0i}=g'(X_i^{\T}\beta_0)/V\{g(X_i^{\T}\beta_0)\}.\nn
 \ee
   $\hfill\Box$
\begin{la}
\label{mean_variance_gt}
The expectations and variances of $A_n$ and $U_n$ are respectively
\be \mu_{\sa}
=\mathrm{tr}\{\Sigma_{\beta}(\beta_0)+\Xi_{\beta,\beta_0}\}, \quad
\mu_{\su}=(n-1)\Delta_{\beta,\beta_{0}}^{\T}\Delta_{\beta,\beta_{0}},
\nn\ee \be \sigma_{\sa}^2 =
{n^{-1}}\bigl[E\bigl\{\epsilon_0^4\psi^4_0(X^\T
X)^2\bigr\}-E^2\bigl\{\epsilon_0^2\psi^2_0(X^\T X)\bigr\}\bigr]\quad
\hbox{and} \quad \nn\ee \be
\sigma_{\su}^2={4(n-2)(1-n^{-1})}\xi_{1}+{2(1-n^{-1})}\xi_{2} \nn\ee
where $\xi_{1} =
\Delta_{\beta,\beta_{0}}^{\T}\{\Sigma_{\beta}(\beta_{0})+
\Xi_{\beta,\beta_{0}}\}\Delta_{\beta,\beta_{0}} -
(\Delta_{\beta,\beta_{0}}^{\T}\Delta_{\beta,\beta_{0}})^{2}$ and~
$\xi_{2} =
\mathrm{tr}\{\Sigma_{\beta}(\beta_{0})+\Xi_{\beta,\beta_{0}}\}^2-(\Delta_{\beta,\beta_{0}}^{\T}\Delta_{\beta,\beta_{0}})^{2}.$
\end{la}

\bigskip

\begin{la}
\label{variance_local} Under Assumptions \ref{as1}-\ref{as4} and the
``local" alternatives $\mathscr{L}_{\beta}$,
\[
\sigma^2_{\su}=2
\mathrm{tr}\{\Sigma_{\beta}(\beta_0)+\Xi_{\beta,\beta_0}\}^2\{1+o(1)\}
\quad  \mbox{as $n\to\infty$.}
\]
\end{la}

\bigskip

\begin{la}
\label{pn1} Under Assumptions \ref{as1}-\ref{as4} and the ``local"
alternatives  $\mathscr{L}_{\beta}$,
 \be
\frac{\widehat{\mathrm{tr}\{\Sigma^2_{\beta_0}(\beta_0)\}}}{\mathrm{tr}\{\Sigma_{\beta}(\beta_0)+\Xi_{\beta,\beta_0}\}^2}\to
1\quad \text{and}\quad
\frac{\widehat{\mathrm{tr}\{\Sigma_{\beta_0}(\beta_0)\}}}{\mathrm{tr}\{\Sigma_{\beta}(\beta_0)+\Xi_{\beta,\beta_0}\}}\to
1 \nn \ee in probability as $n \to \infty$.
\end{la}

\vskip 1cm

\noindent {In the following, we  provide technical proofs for the
main results in Section 4 first, since they are used to establish
the results in Section 3.  The results in Section 5 are given the
last. }

\vskip 1cm
\noindent\textbf{Proof of Theorem \ref{tm1}}
\smallskip

\noindent Define
$\sigma^2_{n}=2\mathrm{tr}\{\Sigma_{\beta}(\beta_0)+\Xi_{\beta,\beta_0}\}^2$.
Notice that
$$U_n-(n-1)\Delta_{\beta,\beta_0}^{\T}\Delta_{\beta,\beta_0}=V_{n1}+V_{n2}\quad \text{where}$$
\[
V_{n1}=\frac{1}{n}\sum\limits_{i\neq
j}^{n}(\Delta_{\beta,\beta_0}^{\T}\epsilon_{0j}\psi_{0j}X_j+\Delta_{\beta,\beta_0}^{\T}\epsilon_{0i}\psi_{0i}X_i-2\Delta_{\beta,\beta_0}^{\T}\Delta_{\beta,\beta_0})\quad
\text{and}\quad
\]
\[
V_{n2}=\frac{1}{n}\sum\limits_{i\neq
j}^{n}(\epsilon_{0i}\psi_{0i}X_i-\Delta_{\beta,\beta_0})^{\T}(\epsilon_{0j}\psi_{0j}X_j-\Delta_{\beta,\beta_0}).
\]
As $E(V_{n1})=0$ and from the Hoeffding decomposition in the proof
of Lemma \ref{mean_variance_gt}, under the ``local" alternatives
$\mathscr{L}_{\beta}$,
\be\text{var}(V_{n1})=o(\sigma^2_n)\quad\text{and}\quad
V_{n1}=o_p(\sigma_n).\nn\ee We use the martingale central limit
theorem to show the asymptotic normality of $V_{n2}$. Let
\[
 Z_{n,i}=\frac{2}{n\sigma_{n}}\sum_{j=1}^{i-1}(\epsilon_{0i}\psi_{0i}X_i-\Delta_{\beta,\beta_0})^{\T}(\epsilon_{0j}\psi_{0j}X_j-\Delta_{\beta,\beta_0})\quad \mbox{for}~i\geq
 2,~\mbox{and}
\]
$T_{n,k}=\sum_{i=2}^{k}Z_{n,i}.$ Then $
T_{n,n}=\sum_{i=2}^{n}Z_{n,i}={V_{n2}}/{\sigma_{n}}.$

\medskip

Let $\mathscr{F}_{k}=\sigma \left\{\left(\begin{smallmatrix}
X_{1}\\\epsilon _{1}
\end{smallmatrix}\right),\ldots ,\left(\begin{smallmatrix}
X_{k}\\\epsilon _{k}
\end{smallmatrix}\right) \right\}$ be the $\sigma$-fields generated by $\bigl(\begin{smallmatrix}
X_{i}\\\epsilon _{i}
\end{smallmatrix}\bigr)$ for $i=1,\ldots ,k.$ It can be verified that $T_{n,k}$ is a  martingale.
For $i=2,\ldots, n$, let $v_{n,i}=E( Z_{n,i}^{2}|
\mathscr{F}_{i-1})$ and $v_{n}=\sum_{i=2}^{n}v_{n,i}$.

\medskip

From \cite{Hall_Heyde_1980}, in order to show the asymptotic
normality of $V_{n2}$, we need to verify the following two
conditions: \be \label{A11}
 v_{n} \to  1\quad \mbox{in probability as $n\to \infty$};
 \ee
\be \label{A12} \text{for any}~\eta > 0,~\sum_{i=2}^{n}E\{
Z_{n,i}^{2}I(| Z_{n,i} |>\eta ) \} \rightarrow 0~~\text{as $n\to
\infty.$} \ee We first establish (\ref{A11}). For $i=2, \dots, n,$
\[
\begin{split}
v_{n,i}
=&\frac{4}{n^2\sigma_{n}^2}\bigg[\sum_{j=1}^{i-1}(\epsilon_{0j}\psi_{0j}X_j-\Delta_{\beta,\beta_0})^{\T}\{\Sigma_{\beta}(\beta_0)+\Xi_{\beta,\beta_0}-\Delta_{\beta,\beta_0}\Delta_{\beta,\beta_0}^{\T}\}(\epsilon_{0j}\psi_{0j}X_j-\Delta_{\beta,\beta_0})\\&+\sum_{j_1\neq
j_2}^{i-1}(\epsilon_{0j_1}\psi_{0j_1}X_{j_1}-\Delta_{\beta,\beta_0})^{\T}\{\Sigma_{\beta}(\beta_0)+\Xi_{\beta,\beta_0}-\Delta_{\beta,\beta_0}\Delta_{\beta,\beta_0}^{\T}\}(\epsilon_{0j_2}\psi_{0j_2}X_{j_2}-\Delta_{\beta,\beta_0})\bigg].
\end{split}
\]
Then \be v_{n}=\sum_{i=2}^{n}v_{n,i}=C_{1}+C_{2}\nn
\quad\text{where} \ee
\[
C_1=\frac{4}{n^2\sigma_{n}^2}\sum_{j=1}^{n-1}\bigl[(n-j)(\epsilon_{0j}\psi_{0j}X_j-\Delta_{\beta,\beta_0})^{\T}\{\Sigma_{\beta}(\beta_0)+\Xi_{\beta,\beta_0}-\Delta_{\beta,\beta_0}\Delta_{\beta,\beta_0}^{\T}\}(\epsilon_{0j}\psi_{0j}X_j-\Delta_{\beta,\beta_0})\bigr]~~\text{and}
\]
\[
C_2=\frac{8}{n^2\sigma_{n}^2}\sum_{1\leq j_{1}<j_{2}\leq
n-1}\bigl[(n-j_2)(\epsilon_{0j_1}\psi_{0j_1}X_{j_1}-\Delta_{\beta,\beta_0})^{\T}\{\Sigma_{\beta}(\beta_0)+\Xi_{\beta,\beta_0}-\Delta_{\beta,\beta_0}\Delta_{\beta,\beta_0}^{\T}\}(\epsilon_{0j_2}\psi_{0j_2}X_{j_2}-\Delta_{\beta,\beta_0})\bigr].
\]
Under the ``local" alternatives $\mathscr{L}_{\beta}$, we have
\[E
\bigl[(\epsilon_{0j}\psi_{0j}X_j-\Delta_{\beta,\beta_0})^{\T}\{\Sigma_{\beta}(\beta_0)+\Xi_{\beta,\beta_0}-\Delta_{\beta,\beta_0}\Delta_{\beta,\beta_0}^{\T}\}(\epsilon_{0j}\psi_{0j}X_j-\Delta_{\beta,\beta_0})\bigr]=\text{tr}\{\Sigma_{\beta}(\beta_0)+\Xi_{\beta,\beta_0}\}^2\{1+o(1)\}.
\]
Thus $E(C_1)={1+o(1)}.$ Similar to the proof of Lemma A.3 in
\cite{Chen_Guo_2014},
\[
\begin{split}
\text{var}(C_{1})
&= \frac{16}{n^4\sigma_{n}^4}\sum_{j=1}^{n-1}E\bigl[(n-j)(\epsilon_{0j}\psi_{0j}X_j-\Delta_{\beta,\beta_0})^{\T}\{\Sigma_{\beta}(\beta_0)+\Xi_{\beta,\beta_0}-\Delta_{\beta,\beta_0}\Delta_{\beta,\beta_0}^{\T}\}(\epsilon_{0j}\psi_{0j}X_j-\Delta_{\beta,\beta_0})\bigr]^{2}\\
&\leq \frac{16}{n^4\sigma_{n}^4}\sum_{j=1}^{n-1} ( n-j
)^{2}O[\text{tr}^2\{\Sigma_{\beta}(\beta_0)+\Xi_{\beta,\beta_0}\}^2]=O(n^{-1}).
\end{split}
\]
Therefore  $C_{1}\to 1$ in probability. For $C_2$, we note that
$E(C_{2})=0$ and
\[
\begin{split}
\text{var}(C_{2})
&=\frac{64}{n^4\sigma_{n}^4}\sum_{1\leq j_{1}<j_{2}\leq n-1}( n-j_{2} )^{2}\text{tr}\{\Sigma_{\beta}(\beta_0)+\Xi_{\beta,\beta_0}-\Delta_{\beta,\beta_0}\Delta_{\beta,\beta_0}^{\T}\}^4=o(1).\\
\end{split}
\]
Thus, $C_{2}\to 0$ in probability. Hence, (\ref{A11}) holds.

Next, we verify (\ref{A12}). Notice that for any $\eta > 0$, \be
\sum_{i=2}^{n}E\{ Z_{n,i}^{2}I( | Z_{n,i} |>\eta )
\}\leq\frac{1}{\eta ^{2}}\sum_{i=2}^{n}E(Z_{n,i}^{4} )\quad
\text{and}\nn \ee
\[
\sum_{i=2}^{n}E(Z_{n,i}^{4})=\frac{16}{n^4\sigma_{n}^4}\sum_{i=2}^{n}E\bigg\{
\sum_{j=1}^{i-1}(\epsilon_{0i}\psi_{0i}X_i-\Delta_{\beta,\beta_0})^{\T}(\epsilon_{0j}\psi_{0j}X_j-\Delta_{\beta,\beta_0})\bigg\}^{4}=P_{1}
+ P_{2}
\]
where
 \be
P_{1}=\frac{16}{n^4\sigma_{n}^4}\sum_{i=2}^{n}\sum_{j=1}^{i-1}E\{(\epsilon_{0i}\psi_{0i}X_i-\Delta_{\beta,\beta_0})^{\T}(\epsilon_{0j}\psi_{0j}X_j-\Delta_{\beta,\beta_0})\}^{4}~~\text{and}~~\nonumber
\ee \be
P_{2}=\frac{16}{n^4\sigma_{n}^4}\sum_{i=2}^{n}\sum_{j_{1}\neq
j_{2}}^{i-1}E[\{
(\epsilon_{0i}\psi_{0i}X_i-\Delta_{\beta,\beta_0})^{\T}(\epsilon_{0j_1}\psi_{0j_1}X_{j_1}-\Delta_{\beta,\beta_0})
(\epsilon_{0i}\psi_{0i}X_i-\Delta_{\beta,\beta_0})^{\T}(\epsilon_{0j_2}\psi_{0j_2}X_{j_2}-\Delta_{\beta,\beta_0})\}^{2}
].\nonumber \ee By  Lemma A.3 in \cite{Chen_Guo_2014} and the
Cauchy-Schwartz inequality, the orders of $P_1$ and $P_2$ are
respectively  \be P_{1}=O(n^{-2})\quad\text{and}\quad
P_2=O(n^{-1}).\nn \ee Then, we obtain
$\sum_{i=2}^{n}E(Z_{n,i}^{4})=o(1)$ and the desired asymptotic
normality of $U_{n}$. \ov




\vskip 1cm

\noindent \textbf{Proof of Theorem \ref{tm2}} \smallskip

\noindent We first show that under the ``fixed" alternatives
$\mathscr{L}_{\beta}^{\mathrm{F}}$,
\be \label{fixednorm}
\frac{U_{n}-(n-1)\Delta_{\beta,\beta_{0}}^{\T}\Delta_{\beta,\beta_{0}}}{\{2\mathrm{tr}(\Xi_{\beta,\beta_0}^2)\}^{1/2}}
\to N(0,1)\quad \mbox{in distribution  as $n\to\infty$}. \ee
Similar to the  proof of Theorem \ref{tm1}, \be \label{la9.1}
U_{n}-(n-1)\Delta_{\beta,\beta_{0}}^{\T}\Delta_{\beta,\beta_{0}}=V_{n1}+V_{n2}+V_{n3}+V_{n4}\quad
\ee where
\[V_{n1}=\frac{1}{n}\sum_{i \neq j}^{n}\epsilon _{i}\epsilon _{j}\psi_{0i}\psi_{0j}X_{i}^{\T}X_{j},
\]
\[
V_{n2}=\frac{1}{n}\sum_{i \neq j}^{n}\bigl[\{(g_{i}-g_{0i})\epsilon
_{j}\psi_{0i}\psi_{0j}X_{i}^{\T}X_{j}
 \}+\{(g_{j}-g_{0j})\epsilon _{i}\psi_{0i}\psi_{0j}X_{i}^{\T}X_{j}
  \}\bigr],
  \]
    \[
   V_{n3}=\frac{1}{n}\sum_{i \neq j}^{n}\bigl[\Delta_{\beta,\beta_0}^{\T}(g_i-g_{0i})\psi_{0i}X_{i}+\Delta_{\beta,\beta_0}^{\T}(g_j-g_{0j})\psi_{0j}X_{j}-2\Delta_{\beta,\beta_0}^{\T}\Delta_{\beta,\beta_0}
   \bigr]\quad \text{and}
   \]
  \[
  V_{n4}=\frac{1}{n}\sum_{i \neq j}^{n}\bigl[\{(g_i-g_{0i})\psi_{0i}X_{i}-\Delta_{\beta,\beta_0}\}^{\T}\{(g_j-g_{0j})\psi_{0j}X_{j}-\Delta_{\beta,\beta_0}\}  \bigr].
  \]
 Notice that  $V_{ni}$ are statistics with zero mean for $i=1,\cdots,4$. Similar to Lemma
 \ref{mean_variance_gt}, we can show
\[
\text{var}(V_{n1})={2(1-n^{-1})}\text{tr}\{\Sigma^2_{\beta}(\beta_0)\}=o\{\text{tr}(\Xi_{\beta,\beta_0}^2)\};
\]
\[
\text{var}(V_{n2})={4(n-2)(1-n^{-1})}\Delta_{\beta,\beta_0}^{\T}\Sigma_{\beta}(\beta_0)\Delta_{\beta,\beta_0}+{4(1-n^{-1})}\text{tr}\{\Sigma_{\beta}(\beta_0)\Xi_{\beta,\beta_0}\}=o\{\text{tr}(\Xi_{\beta,\beta_0}^2)\};
\]
\[
\text{var}(V_{n3})\leq
{4(n-2)(1-n^{-1})}\Delta_{\beta,\beta_0}^{\T}\Xi_{\beta,\beta_0}\Delta_{\beta,\beta_0}+{4(1-n^{-1})}\Delta_{\beta,\beta_0}^{\T}\Xi_{\beta,\beta_0}\Delta_{\beta,\beta_0}=o\{\text{tr}(\Xi_{\beta,\beta_0}^2)\};
\]
Then
\[
V_{n1}=o_p\{\text{tr}^{{1}/{2}}(\Xi_{\beta,\beta_0}^2)\},~~V_{n2}=o_p\{\text{tr}^{{1}/{2}}(\Xi_{\beta,\beta_0}^2)\}~~\text{and}~~V_{n3}=o_p\{\text{tr}^{{1}/{2}}(\Xi_{\beta,\beta_0}^2)\}.
\]
Applying the same technique  we used in the proof of Theorem
\ref{tm1}, we have
\[
\frac{V_{n4}}{\{2\text{tr}(\Xi_{\beta,\beta_0}^2)\}^{1/2}} \to
N(0,1)\quad \mbox{in distribution as $n \to \infty.$}
\]
Then from the decomposition (\ref{la9.1}), the asymptotic normality
(\ref{fixednorm}) holds. The power expression stated in the theorem
is readily available from Lemma \ref{fixednorm}.   \ov

\vskip 1cm

\noindent \textbf{Proof of Theorem \ref{tmGoeman_local}}

\smallskip

\noindent Note that
\[
\widehat{S}_n=1+\frac{\{(\by-\bmu_0)\circ
\bPsi_0\}^{\T}({\bx}\bx^{\T}-\bd)\{(\by-\bmu_0)\circ
\bPsi_0\}/n}{\{({\by }-\bmu_0)\circ
\bPsi_0\}^{\T}\bd\{(\by-\bmu_0)\circ \bPsi_0\}/n}=
1+\frac{U_n}{A_n}.
\]
Let
$\mu_\sb=E(U_n)=(n-1)\Delta_{\beta,\beta_0}^{\T}\Delta_{\beta,\beta_0}$
and $
\mu_\sa=E(A_n)=\text{tr}\{\Sigma_{\beta}(\beta_0)+\Xi_{\beta,\beta_0}\}.$
From the Taylor expansion, \be
\begin{split}
\widehat{S}_n=&1+\frac{\mu_\sb+(U_n-\mu_\sb)}{\mu_\sa(1+\frac{A_n-\mu_\sa}{\mu_\sa})}\\=&
1+\mu_\sa^{-1}\biggl\{1-\frac{A_n-\mu_\sa}{\mu_\sa}+(\frac{A_n-\mu_\sa}{\mu_\sa})^2+\cdots \biggr\}\{\mu_\sb+(U_n-\mu_\sb)\}\\
=&
1+\mu_\sa^{-1}\mu_\sb-\mu_\sa^{-1}\mu_\sb(\frac{A_n-\mu_\sa}{\mu_\sa})+\mu_\sa^{-1}(U_n-\mu_\sb)+\mu_\sa^{-1}\mu_\sb(\frac{A_n-\mu_\sa}{\mu_\sa})^2+\cdots.
\label{Atm1_2}
\end{split}
\ee
Under the ``local" alternatives $\mathscr{L}_{\beta}$, from Lemma
A.3 in \cite{Chen_Guo_2014},
$$\sigma_\sa^2\leq n^{-1}E\{\epsilon_0^4\psi_0^4(X^\T
X)^2\}=O\{n^{-1}\text{tr}^2(\Sigma_\sx)\}, \quad
\sigma_\sa^2/\mu_\sa^2=O(n^{-1})\quad \text{and}$$ $$
{\sigma_\sb^{-1}}{(U_n-\mu_\sb)}\to N(0,1) ~~\mbox{in distribution
as $n \to \infty$.}
$$
Observe that
\[
\text{var}\{\mu_\sa^{-2}\mu_{\su}(A_n-\mu_\sa)\}=\sigma_\sa^2\mu_{\su}^2\mu_\sa^{-4}
\quad \text{and}\quad
\text{var}\{\mu_\sa^{-1}(U_n-\mu_\sb)\}={\sigma_\sb^2}\mu_\sa^{-2}.
\]
%
From the fact that
\[
({\sigma_\sa^2\mu_{\su}^2})/({\mu_\sa^2\sigma_\su^2})=O\bigl({n(\Delta_{\beta,\beta_{0}}^{\T}\Delta_{\beta,\beta_{0}})^2}/{\text{tr}\{\Sigma_{\beta}(\beta_0)+\Xi_{\beta,\beta_0}\}^2}\bigr)=o(1),
\]
we have
$\mu_\sa^{-2}\mu_{\su}(A_n-\mu_\sa)=o_p({\sigma_\sb}{\mu_\sa^{-1}})$.

Regarding the higher order terms in the expansion
(\ref{Atm1_2}), for $k\geq 1$,
\[
(\frac{A_n-\mu_\sa}{\mu_\sa})^k(\frac{U_n-\mu_\sb}{\mu_\sa})=(\frac{A_n-\mu_\sa}{\sigma_\sa})^k(\frac{U_n-\mu_\sb}{\sigma_\sb})\frac{\sigma_\sa^k}{\mu_\sa^k}\frac{\sigma_\sb}{\mu_\sa}=O_p(n^{-\frac{k}{2}}\frac{\sigma_\sb}{\mu_\sa}).
\]
Note that under the ``local" alternatives $\mathscr{L}_{\beta}$,
\[
\frac{\sigma_\sa^2}{\mu^2_\sa}\frac{\mu_\sb}{\sigma_\sb}=O(n^{-1})\frac{(n-1)\Delta_{\beta,\beta_{0}}^{\T}\Delta_{\beta,\beta_{0}}}{[{2\text{tr}\{\Sigma_{\beta}(\beta_{0})+\Xi_{\beta,\beta_0}\}^2}]^{1/2}}\{1+o(1)\}=o(1).
\]
Hence, for $k\geq 2$,
\[
\mu_\sa^{-1}\mu_\sb(\frac{A_n-\mu_\sa}{\mu_\sa})^k=(\frac{A_n-\mu_\sa}{\sigma_\sa})^k\frac{\sigma_\sa^{k-2}}{\mu_\sa^{k-2}}\frac{\sigma_\sa^2}{\mu^2_\sa}\frac{\mu_\sb}{\sigma_\sb}\frac{\sigma_\sb}{\mu_\sa}=o_p(n^{-\frac{k-2}{2}}\frac{\sigma_\sb}{\mu_\sa})=o_p(\frac{\sigma_\sb}{\mu_\sa}).
\]
Therefore, 
 \be
\widehat{S}_n=1+\mu_{\sa}^{-1}\mu_{\sb}+\mu_\sa^{-1}(U_n-\mu_\sb)+
o_p({\sigma_\sb}{\mu^{-1}_\sa}).
\label{Atm1_3} \ee From the asymptotic normality of $U_n$ and the
Slutsky theorem, we have
\[
{\sigma_\sb^{-1}}{\mu_\sa}(\widehat{S}_n-1-\mu_{\sa}^{-1}\mu_{\sb})\to
N(0,1) ~~\mbox{in distribution as $n \to \infty$}.
\]
\ov



\noindent

\vskip 1cm

\noindent \textbf{Proof of Theorem \ref{tmGoeman_fixed}}
\smallskip

\noindent For brevity,
we define \be
\begin{split}
\sigma^2_{\sg}=&\text{var}\bigl\{
-\mu_{\sb}\mu_\sa^{-2}(A_n-\mu_\sa)+\mu_\sa^{-1}(U_n-\mu_\sb)\bigr\}\\
=&{\sigma_\sb^2}{\mu_\sa^{-2}}\{1+\tau(\tau-2\rho_{\sa,\sb})\}
\end{split}
\nn \ee where
$\tau^2=(\sigma_{\sa}^2\mu_{\sb}^2)/(\sigma_{\sb}^2\mu_{\sa}^2)$ and
$\rho_{\sa,\sb}$ is the correlation coefficient between $A_n$ and
$U_n$. It is straightforward to show that, by Cauchy-Schwartz
inequality,
\[
\begin{split}
\text{cov}(A_n,U_n)
&\leq 2E^{{1}/{2}}\bigl[\epsilon_0^4\psi_0^4(X^\T
X)^2-E^2\{\epsilon_0^2\psi_0^2(X^\T
X)\}\bigr]\bigl[\Delta_{\beta,\beta_0}^{\T}\{\Xi_{\beta,\beta_0}+\Sigma_{\beta}(\beta_0)\}\Delta_{\beta,\beta_0}
\bigr]^{{1}/{2}}.
\end{split}
\]
Notice that under the ``fixed" alternatives
$\mathscr{L}_{\beta}^{\text{F}}$,
\[
E\bigl[\epsilon_0^4\psi_0^4(X^\T X)^2-E^2\{\epsilon_0^2\psi_0^2(X^\T
X)\}\bigr]=n\sigma_{\sa}^2~~\text{and}~~
\Delta_{\beta,\beta_0}^{\T}\{\Xi_{\beta,\beta_0}+\Sigma_{\beta}(\beta_0)\}\Delta_{\beta,\beta_0}=o(n^{-1}\sigma_{\sb}^2).
\]
Therefore
$$\text{cov}(A_n,U_n)=o(\sigma_{\sa}\sigma_{\sb})\quad
\text{and}\quad
 \rho_{\sa,\sb}=o(1).$$Recall that
\[
\tau^2=\frac{\sigma_{\sa}^2\mu_{\sb}^2}{\sigma_{\sb}^2\mu_{\sa}^2}=\frac{n(\Delta_{\beta,\beta_0}^{\T}\Delta_{\beta,\beta_0})^2}{\text{tr}(\Xi_{\beta,\beta_0}^2)}\biggl[
\frac{E\{\epsilon_0^4\psi_0^4(X^\T
X)^2\}}{E^2\{\epsilon_0^2\psi_0^2(X^\T X)\}}-1 \biggr]\{1+o(1)\}.
\]
The condition (\ref{conditionTh3}) implies
\[
\frac{n(\Delta_{\beta,\beta_0}^{\T}\Delta_{\beta,\beta_0})^2}{\text{tr}(\Xi_{\beta,\beta_0}^2)}
\frac{E\{\epsilon_0^4\psi_0^4(X^\T
X)^2\}}{E^2\{\epsilon_0^2\psi_0^2(X^\T X)\}}\asymp 1.
\]
Thus, $\tau^2\asymp 1$  {and}
$\sigma^2_{\sg}={\sigma_\sb^2}{\mu_\sa^{-2}}\{1+\tau^2+o(1)\}.$
Observe that
\[
\frac{\sigma_{\sa}^2}{\mu_{\sa}^2}\leq
\frac{1}{n}\frac{E\{\epsilon_0^4\psi_0^4(X^{\T}X)^2\}}{E^2\{\epsilon_0^2\psi_0^2(X^{\T}X)\}}\asymp
n^{-2\delta}.
\]
Regarding the higher order terms in  (\ref{Atm1_2}), for $k\geq 1$,
\[
(\frac{A_n-\mu_\sa}{\mu_\sa})^k(\frac{U_n-\mu_\sb}{\mu_\sa})=(\frac{A_n-\mu_\sa}{\sigma_\sa})^k(\frac{U_n-\mu_\sb}{\sigma_\sb})\frac{\sigma_\sa^k}{\mu_\sa^k}\frac{\sigma_\sb}{\mu_\sa}=O_p(n^{-k\delta}\frac{\sigma_\sb}{\mu_\sa})=O_p(n^{-k\delta}\sigma_{\sg}).
\]
Notice that under the ``fixed" alternatives
$\mathscr{L}_{\beta}^{\mathrm{F}}$
\[
\frac{\sigma_\sa^2}{\mu^2_\sa}\frac{\mu_\sb}{\sigma_\sb}=O(n^{-2\delta})\frac{n\Delta_{\beta,\beta_{0}}^{\T}\Delta_{\beta,\beta_{0}}}{\{{2\text{tr}(\Xi^2_{\beta,\beta_0})}\}^{1/2}}\{1+o(1)\}=O(n^{-\delta})=o(1).
\]
Then, for $k\geq 2$,
\[
\mu_\sa^{-1}\mu_\sb(\frac{A_n-\mu_\sa}{\mu_\sa})^k=(\frac{A_n-\mu_\sa}{\sigma_\sa})^k(\frac{\sigma_\sa^{k-2}}{\mu_\sa^{k-2}})(\frac{\sigma_\sa^2}{\mu^2_\sa}\frac{\mu_\sb}{\sigma_\sb})\frac{\sigma_\sb}{\mu_\sa}=o_p(n^{-(k-2)\delta}\frac{\sigma_\sb}{\mu_\sa})=o_p(\sigma_{\sg}).
\]
It follows that, 
\be
\widehat{S}_n=1+\mu_\sa^{-1}\mu_\sb-\mu_\sb\mu_\sa^{-2}(A_n-\mu_\sa)+\mu_\sa^{-1}(U_n-\mu_\sb)+o_p(\sigma_{\sg}).
\nn \ee From the joint asymptotic normalities of $A_n-\mu_\sa$ and
$U_n-\mu_\sb$, we have
\[
\sigma_{\sg}^{-1}(\widehat{S}_n-1-\mu_\sa^{-1}\mu_\sb)\to
N(0,1)~~\text{in distribution,}
\]
where
$\sigma_{\sg}^2={\sigma_\sb^2}{\mu_\sa^{-2}}\{1+\tau^2+o(1)\}$.

Analogous to the proof of Lemma  \ref{pn1}, we can show that under
the ``fixed" alternatives $\mathscr{L}_{\beta}^{\mathrm{F}}$,
\[
\sigma_{\sg}^{-1}\bigl[{2\widehat{\text{tr}\{\Sigma^2_{\beta}(\beta_0)\}}/\widehat{\text{tr}^2\{\Sigma_{\beta}(\beta_0)\}}}\bigr]^{1/2}\to
\frac{1}{(1+\tau^2)^{1/2}}\quad \text{in probability as $n \to
\infty$}.
\]
Together with the asymptotic normality of $\widehat{S}_n$, we
complete the proof.
\ov


\vskip 1cm

\noindent \textbf{Proof of Theorem \ref{tm3}}
\smallskip

\noindent Since the proof of the theorem is long, we divide it into
two lemmas.
\begin{la}
\label{lemma5} Suppose Assumptions \ref{as1}, \ref{as3},
\ref{as5}-\ref{as8} hold, then under the $H_{01}$, \be
\frac{\widetilde{U}_n}{[{2\mathrm{tr}\{\Sigma^{\ss2}_{\beta}({\beta}_0^{\ast})\}^2}]^{1/2}}\to
N(0,1)\nn \ee in distribution as $n\to \infty$.
\end{la}
\begin{la}
\label{lemma6} Suppose Assumptions \ref{as1}, \ref{as3},
\ref{as5}-\ref{as8} hold, under the ``local" alternatives
$\mathscr{L}_{{\beta}^{\ss2}}$,
  \be
\frac{\widetilde{U}_n-n\Delta_{\beta,{\beta}_0^{\ast}}^{\ss2\T}\Delta_{\beta,{\beta}_0^{\ast}}^{\ss2}}{[{2\mathrm{tr}\{\Sigma^{\ss2}_{\beta}({\beta}_0^{\ast})\}^2}]^{1/2}}\to
N(0,1)\nn \ee in distribution as  $n \to \infty$.
\end{la}
\vskip 1cm

\noindent \textbf{Proof of Lemma \ref{lemma5}}
\smallskip

\noindent
 Recall that $\hat{\beta}^{\s1}_0$ is the maximum
quasi-likelihood estimator of $\beta^{\s1}$ under $H_{01}$ and
$\beta^{*\s1}=\beta^{\s1}$. For notational convenience, we let
$\hat{\beta}_0=(\hat{\beta}_0^{\s1\T},\beta_0^{\ss2\T})^{\T}$,
${\beta}_0=({\beta}^{\s1\T},\beta_0^{\ss2\T})^{\T}$ and
\[\hat{\mu}_{0i}=g(X_{i}^{\T}\hat{\beta}_0),\quad {\mu}_{0i}=g(X_{i}^{\T}{\beta}_0),\quad \hat{\psi}_{0i}=g'(X_{i}^{\T}\hat{\beta}_0)/V\{g(X_{i}^{\T}\hat{\beta}_0)\},
\quad
{\psi}_{0i}=g'(X_{i}^{\T}{\beta}_0)/V\{g(X_{i}^{\T}{\beta}_0)\};\]
\[
\widehat{\bmu}_0=(\hat{\mu}_{01}, \ldots ,
\hat{\mu}_{0n})^\T,~~\bmu_0=(\mu_{01}, \ldots ,
\mu_{0n})^\T,~~\widehat{\bPsi}_0=(\hat{\psi}_{01}, \ldots,
\hat{\psi}_{0n})^{\T},~~\bPsi_0=(\psi_{01}, \ldots, \psi_{0n})^{\T};
\]
\[
g'_{0i}={\partial g(t)/\partial t}~|~_{t=X_i^{\T}\beta_0},\quad
\psi'_{0i}={\partial \psi\{g(t)\}/\partial
g}~|~_{t=X_i^{\T}\beta_0}.
\]
Define $\widehat{{\bd}}=(\by-\widehat{\bmu}_0)\circ
\widehat{\bPsi}_0=\{(Y_1-\hat{\mu}_{01})\hat{\psi}_{01}, \dots,
(Y_n-\hat{\mu}_{0n})\hat{\psi}_{0n}\}^{\T}$. Then we can write
$\widetilde{U}_{n}$ as \[
\begin{split}
\widetilde{U}_{n}=&{n^{-1}}\{(\by-\widehat{\bmu}_0)\circ
\widehat{\bPsi}_0\}^{\T}(\bx^{\ss2}{\bx}^{\ss2\T}-\Omega)\{(\by-\widehat{\bmu}_0)\circ
\widehat{\bPsi}_0\}={n^{-1}}\widehat{\bd}^{\T}(\bx^{\ss2}{\bx}^{\ss2\T}-\Omega)\widehat{\bd}
\end{split}
\] where $\Omega$ is the diagonal matrix
with diagonal elements being those of $\bx^{\ss2}{\bx}^{\ss2\T}$.

Following the approach in \cite{le Cessie_van Houwelingen_1991_Bio},
we
have
 \be
 \widehat{\bd}=[\bi_n+(\bw_2-\bw_1)\bx^{\s1}\{{\bi}(\beta^{\s1})\}^{-1}\bx^{\s1\T}]\bd\label{expanD}
 \ee
where $\bi_n$ is the $n\times n$ identity matrix, $\bw_1$ and
$\bw_2$ are two diagonal matrices defined as
\[
\bw_1
=\text{diag}\{\psi_{01}^2E(\epsilon_{01}^2| X_1), \ldots,
\psi_{0n}^2E(\epsilon_{0n}^2|X_n)\}\quad \text{and}\]\[
\bw_2=\text{diag}\{ \psi'_{01}g'_{01}(Y_1-g_{01}),\dots,
\psi'_{0n}g'_{0n}(Y_n-g_{0n})\}.
\]
Moreover, $\bd=(\by-\bmu_0)\circ \bPsi_0$ and ${\bi}(\beta^{\s1})$
is a $p_1 \times p_1$  matrix given by $
{\bi}(\beta^{\s1})=\bx^{\s1\T}{\bw}_1\bx^{\s1}.$

In order to simplify the notations, let
$$\ba=(\bx^{\ss2}{\bx}^{\ss2\T}-{\Omega})=(a_{ij})_{n\times
n}\quad\text{and}\quad\bb=\bx^{\s1}\{{\bi}(\beta^{\s1})\}^{-1}\bx^{\s1\T}=(b_{ij})_{n\times
n}.$$ Therefore, by (\ref{expanD}), we can decompose the statistic
$\widetilde{U}_{n}$ as
 \[
 \begin{split}
\widetilde{U}_{n}
   =&~n^{-1}\bd^{\T}\ba\bd+n^{-1}{\bd}^{\T}{\bb}\bw_1\ba\bw_1\bb\bd+n^{-1}{\bd}^{\T}{\bb}\bw_2\ba\bw_2\bb\bd\\
   &+2n^{-1}{\bd}^{\T}\ba\bw_2\bb\bd-2n^{-1}\bd^{\T}\bb\bw_1\ba\bw_2\bb\bd-2n^{-1}{\bd}^{\T}\ba\bw_1\bb\bd\\
    =&~T_{n1}+T_{n2}+T_{n3}+2T_{n4}-2T_{n5}-2T_{n6},\quad
    \text{say}.
\end{split}
\]
 Notice that under $H_{01}$, by the properties of conditional expectation and  Assumption  3.3,
 \be
 E(\epsilon_{0i}| X_i^{\ss2})=E\{E(\epsilon_{0i}| X_i)| X_i^{\ss2}\}=0, \qquad   E(\epsilon_{0i}^2| X_i^{\ss2})=E\{E(\epsilon_{0i}^2| X_i)| X_i^{\ss2}\}\geq K_1 \nn
\ee \be\text{and}\quad
 E(\epsilon_{0i}^8| X_i^{\ss2})=E\{E(\epsilon_{0i}^8| X_i)| X_i^{\ss2}\}\leq K_2.\nn
\ee From Assumption \ref{as1}, we can partition $X_i$ and $\Gamma$
respectively as \be X_i=\begin{pmatrix}
X^{\s1}_{i_{p_1\times 1}}\\
                       \\
X^{\ss2}_{i_{p_2\times 1}}
\end{pmatrix}=\begin{pmatrix}
\Gamma_1Z_i\\
\Gamma_2Z_i
\end{pmatrix} \quad\text{and}\quad \Gamma=\begin{pmatrix}
\Gamma_{1_{p_1\times m}}\\
\Gamma_{2_{p_2\times m}}
\end{pmatrix}.
\nn \ee Furthermore, we have
$\Sigma_{\sx^{\ss2}}=\Gamma_2\Gamma_2^{\T}$. This indicates that the
model in Assumption \ref{as1} still holds for $X_i^{\ss2}$, except
we replace $\Sigma_\sx$ as $\Sigma_{\sx^{\ss2}}$, $\Gamma$ as
$\Gamma_2$.

Under the null hypothesis,
by Assumptions \ref{as1}, \ref{as3}, \ref{as5} and \ref{as8}, the
same technique used in the proof of Theorem \ref{tm1} leads to  \be
\frac{T_{n1}}{[{2\text{tr}\{\Sigma^{\ss2}_{\beta}({\beta}_0^{\ast})\}^2}]^{1/2}}\to
N(0,1)~~ \mbox{in distribution as $n \to \infty.$}\nn \ee
In the following proofs, we denote all the constants by $C$ which
may vary from place to place. 

Observe that \be |T_{n2}|\leq
n^{-1}|{\bd}^{\T}{\bb}\bw_1\ba\bw_1\bb\bd|\leq
n^{-1}(|\lambda_{\max}(\ba)|\wedge
|\lambda_{\min}(\ba)|)\bd^{\T}{\bb}\bw_1^2\bb\bd.\label{A17}\ee
By the method of \cite{Lan_Wang_Tsai_2013}, we can show that
\be \label{tm4_0}
 |\lambda_{\max}({\bf
\ba})|=O_p\bigl(n^{{3}/{4}}\text{tr}^{{1}/{2}}[\{\Sigma^{\ss2}_{\beta}({\beta}_0^{\ast})\}^2]\bigr)
\ee
and the same order holds for $|\lambda_{\min}(\ba)|$.

 From the independence among the observations and
$E(\epsilon_{0i}|X_i)=0$, \be
E(\bd^{\T}\bb\bw_1^2\bb{\bd})=\sum\limits_{i=1}^{n}\sum\limits_{k=1}^{n}E(b_{ik}^2\psi_{0i}^2\psi_{0k}^2g_{0i}'^2\epsilon_{0k}^2)\leq
C\sum\limits_{i=1}^{n}\sum\limits_{k=1}^{n}E(b_{ik}^2)=CE\{\text{tr}({\bb}^2)\}.\label{tm4_1}
\ee Notice that from Assumptions \ref{as3} and \ref{as8}, we have
\be
\bi(\beta^{\s1})=\bx^{\s1\T}\bw_1{\bx}^{\s1}=\sum\limits_{i=1}^{n}\{\psi_{0i}^2E(\epsilon_{0i}^2|
X_i)X_i^{\s1}X_i^{\s1\T}\}\geq K_1c_1
\sum\limits_{i=1}^{n}X_i^{\s1}X_i^{\s1\T}=K_1c_1{\bx}^{\s1\T}\bx^{\s1}.\nn
\ee Together with the matrix inequality from \cite{Seber_2008}, we
get \bea \label{A2}
E\{\text{tr}(\bb^2)\}=E\bigl(\text{tr}[\bx^{\s1}\{{\bi}(\beta^{\s1})\}^{-1}\bx^{\s1\T}\bx^{\s1}\{{\bi}(\beta^{\s1})\}^{-1}\bx^{\s1\T}]\bigr)
\leq K^{-2}_1c^{-2}_1p_1.
 \eea
Thus, the order of $T_{n2}$ is \be
T_{n2}=O(n^{-1})O_p\bigl(n^{{3}/{4}}\text{tr}^{{1}/{2}}[\{\Sigma^{\ss2}_{\beta}({\beta}_0^{\ast})\}^2]\bigr)O_p(p_1)=o_p\bigl(\text{tr}^{{1}/{2}}[\{\Sigma^{\ss2}_{\beta}({\beta}_0^{\ast})\}^2]\bigr).\nn
\ee
Applying the same technique, we can show
\[
T_{n3}=o_p\bigl(\text{tr}^{{1}/{2}}[\{\Sigma^{\ss2}_{\beta}({\beta}_0^{\ast})\}^2]\bigr).
\]
For the order of $T_{n4}$,
to simplify the notations, we define
\be
\bd^{\T}\ba{\bw}_2=\bigg(\sum\limits_{k=1}^{n}a_{k1}\psi'_{01}g'_{01}\psi_{0k}\epsilon_{0k}\epsilon_{01},
\dots,
\sum\limits_{k=1}^{n}a_{kn}\psi'_{0n}g'_{0n}\psi_{0k}\epsilon_{0k}\epsilon_{0n}\bigg)=(c_{01},
\dots, c_{0n}),
\quad \mbox{say;}\nn \ee 
\be \bb{\bd}=\bigg(
\sum\limits_{k=1}^{n}b_{1k}\psi_{0k}\epsilon_{0k},\dots,\sum\limits_{k=1}^{n}b_{nk}\psi_{0k}\epsilon_{0k}\bigg)^{\T}
= (f_{01}, \dots, f_{0n})^{\T}, \quad \mbox{say.}\nn \ee
Then \be \label{A18}
E(\bd^{\T}\ba\bw_2\bb{\bd})^2=\sum\limits_{i=1}^{n}E(c_{0i}^2f_{0i}^2)+\sum\limits_{i_1\neq
i_2}^{n}E(c_{0i_1}c_{0i_2}f_{0i_1}f_{0i_2}). \ee We can write
 \be
\sum\limits_{i=1}^{n}E(c_{0i}^2f_{0i}^2)=T_{41}+2T_{42}+2T_{43}
\quad  \label{A19} \ee \text{where} \be
T_{41}=\sum\limits_{i=1}^{n}\sum\limits_{k_1=1}^{n}\sum\limits_{k_2=1}^{n}E(a_{k_1i}^2b_{ik_2}^2\psi'^2_{0i}g'^2_{0i}\psi_{0k_1}^2\psi_{0k_2}^2\epsilon_{0k_1}^2\epsilon_{0k_2}^2\epsilon_{0i}^2),
\nn\ee \be
T_{42}=\sum\limits_{i=1}^{n}\sum\limits_{k=1}^{n}E(a_{ki}^2b_{ii}b_{ik}\psi_{0k}^3\psi_{0i}\psi'^2_{0i}g'^2_{0i}\epsilon_{0k}^3\epsilon_{0i}^3)\quad\text{and}
\nn\ee \be T_{43}=\sum\limits_{i=1}^{n}\sum\limits_{k_1\neq
k_2}^{n}E(a_{k_1i}a_{k_2i}b_{ik_1}b_{ik_2}\psi_{0k_1}^2\psi_{0k_2}^2\psi'^2_{0i}g'^2_{0i}\epsilon_{0k_1}^2\epsilon_{0k_2}^2\epsilon_{0i}^2).\nn
\ee
Notice that
 \be T_{41} \leq
C\sum\limits_{i=1}^{n}\sum\limits_{k_1=1}^{n}\sum\limits_{k_2=1}^{n}E(a_{k_1i}^2b_{ik_2}^2).\nn
\ee From the Cauchy-Schwartz inequality, \be \label{A20}
\sum\limits_{i=1}^{n}\sum\limits_{k_1=1}^{n}\sum\limits_{k_2=1}^{n}E(a_{k_1i}^2b_{ik_2}^2)\leq
E^{{1}/{2}}\bigg\{\sum\limits_{i=1}^{n}\bigg(\sum\limits_{k_1=1}^{n}a_{k_1i}^2\bigg)^2\bigg\}E^{{1}/{2}}\bigg\{\sum\limits_{i=1}^{n}\bigg(\sum\limits_{k_2=1}^{n}b_{ik_2}^2\bigg)^2\bigg\}.
\ee
Recall that $a_{ii}=0$, from Lemma A.2 in \cite{Chen_Guo_2014}, we
have \be \label{A21}
E\bigg\{\sum\limits_{i=1}^{n}\bigg(\sum\limits_{k_1=1}^{n}a_{k_1i}^2\bigg)^2\bigg\}
=O\bigl(n^3\text{tr}^2[\{\Sigma^{\ss2}_{\beta}({\beta}_0^{\ast})\}^2]\bigr).
\ee
On the other hand, $\text{rank}(\bb)\leq p_1$ and employing the same
technique as we used in the derivation of (\ref{A2}), we have \be
\label{A22}
\begin{split}
E\bigg\{\sum\limits_{i=1}^{n}\bigg(\sum\limits_{k_2=1}^{n}b_{ik_2}^2\bigg)^2\bigg\}
&\leq E\{\text{tr}^2(\bb^2)\}\leq
E\{\text{rank}(\bb)\text{tr}(\bb^4)\}=O(p_1^2).
\end{split}
\ee
Combining (\ref{A20}), (\ref{A21}) and (\ref{A22}),
 \be
T_{41}=O(n^{{3}/{2}})O_p(p_1)O_p\bigl(\text{tr}[\{\Sigma^{\ss2}_{\beta}({\beta}_0^{\ast})\}^2]\bigr)=o_p\bigl(n^{{7}/{4}}\text{tr}[\{\Sigma^{\ss2}_{\beta}({\beta}_0^{\ast})\}^2]\bigr).\nn
\ee Applying the same method to that for $T_{41}$, we can show \be
T_{42}=o_p\bigl(n^{{7}/{4}}\text{tr}[\{\Sigma^{\ss2}_{\beta}({\beta}_0^{\ast})\}^2]\bigr)\quad\text{and}\quad
T_{43}=o_p\bigl(n^{{7}/{4}}\text{tr}[\{\Sigma^{\ss2}_{\beta}({\beta}_0^{\ast})\}^2]\bigr).
\nn \ee Then from (\ref{A19}), we have
\be
\sum\limits_{i=1}^{n}E(c_{0i}^2f_{0i}^2)=o\bigl(n^{{7}/{4}}\text{tr}[\{\Sigma^{\ss2}_{\beta}({\beta}_0^{\ast})\}^2]\bigr).
\label{A25} \ee
Derivations given in \cite{Chen_Guo_2014} show \be \label{A28}
\sum\limits_{i_1\neq
i_2}^{n}E(c_{0i_1}c_{0i_2}f_{0i_1}f_{0i_2})=o\bigl(n^{2}\text{tr}[\{\Sigma^{\ss2}_{\beta}({\beta}_0^{\ast})\}^2]\bigr).
\ee Combining (\ref{A18}), (\ref{A25}) and (\ref{A28}), we have
$$T_{n4}=o_p\bigl(\text{tr}^{{1}/{2}}[\{\Sigma^{\ss2}_{\beta}({\beta}_0^{\ast})\}^2]\bigr).$$
From (\ref{tm4_0}),   (\ref{tm4_1}) and (\ref{A2}), we can show
 \be
|T_{n5}|
\leq {n^{-1}}(|\lambda_{\max}(\ba)| \wedge |\lambda_{\min}(\ba)|)
({\bd}^{\T}\bb\bw_1^2\bb\bd)^{1/2}(\bd^{\T}\bb{\bw}_2^2{\bb}\bd)^{1/2}=o_p\bigl(\text{tr}^{{1}/{2}}[\{\Sigma^{\ss2}_{\beta}({\beta}_0^{\ast})\}^2]\bigr).\nn
\ee
 For the order of $T_{n6}$, notice that $\bb$ is a
non-negative matrix, then \be
E|T_{n6}|={n^{-1}}E|{\bd}^{\T}\ba\bw_1\bb\bd|\leq
{n^{-1}}E^{{1}/{2}}(\bd^{\T}\ba\bw_1\bb{\bw}_1\ba\bd)E^{{1}/{2}}(\bd^{\T}\bb\bd).\nn
\ee By the definitions of $\bw_1$ and $\bd$, it is straightforward
to see \be
{\bw}_1=\text{diag}\{\psi_{01}^2E(\epsilon_{01}^2|X_1),\cdots,
\psi_{0n}^2E(\epsilon_{0n}^2|X_n)\}=E(\bd\bd^{\T}|\bx).\nn \ee
Applying some basic matrix inequalities, we have
\[
\begin{split}
E(\bd^{\T}\ba\bw_1\bb\bw_1\ba\bd)&=\text{tr}E(\bw_1\ba\bw_1\bb\bw_1\ba)
          \leq  E^{{1}/{2}}\{\text{tr}(\ba\bw_1\ba\bw_1^2\ba\bw_1\ba)\}E^{{1}/{2}}\{\text{tr}(\bb\bw_1^2\bb)\}
          \end{split}
\]
\[
\begin{split}
\text{and}\quad E\{\text{tr}(\ba\bw_1\ba\bw_1^2\ba\bw_1\ba)\}\leq
CE\{\text{tr}(\ba^4)\}=O\bigl(n^{3}\text{tr}^2[\{\Sigma^{\ss2}_{\beta}({\beta}_0^{\ast})\}^2]\bigr).
                         \end{split}
\] It can be shown that $E\{\text{tr}(\bb\bw_1^2\bb)\}=O(p_1)$ and
$E(\bd^{\T}\bb{\bd})=O(p_1).$ Thus,
\be
T_{n6}=O(n^{-1})O_p\bigl(n^{{3}/{4}}\text{tr}^{{1}/{2}}[\{\Sigma^{\ss2}_{\beta}({\beta}_0^{\ast})\}^2]\bigr)O(p_1^{{1}/{4}})O_p(p_1^{{1}/{2}})=o_p\bigl(\text{tr}^{{1}/{2}}[\{\Sigma^{\ss2}_{\beta}({\beta}_0^{\ast})\}^2]\bigr).\nn
\ee Therefore,
 the asymptotic normality of $T_{n1}$ and the orders of
$T_{n2},\ldots, T_{n6}$ lead to \be
\frac{\widetilde{U}_n}{[{2\text{tr}\{\Sigma^{\ss2}_{\beta}({\beta}_0^{\ast})\}^2}]^{1/2}}\to
N(0,1)\quad \mbox{in distribution as $n \to \infty.$}
 \nn \ee  $\hfill \Box$

\vskip 1cm

\noindent \textbf{Proof of Lemma \ref{lemma6}}

\smallskip

\noindent
 Define
$$\beta_0^{*}=(\beta^{*\s1\T},\beta_0^{\ss2\T})^{\T},\quad
\beta=(\beta^{\s1\T},\beta^{\ss2\T})^{\T},\quad
g_{0i}^{*}=g(X_i^{\T}\beta_0^{*}),\quad g_{i}=g(X_i^{\T}\beta),$$
\[
\psi_{0i}^{*}=g'(X_i^{\T}\beta_0^{*})/V\{g(X_i^{\T}\beta_0^{*})\},\quad
g^{*'}_{0i}={\partial g(t)/\partial t}~|~_{t=X_i^{\T}\beta_0^{*}},
\]
\[
\psi^{*'}_{0i}={\partial \psi\{g(t)\}/\partial
g}~|~_{t=X_i^{\T}\beta^{*}_0}, \quad
\epsilon_i=Y_i-g_i\quad\text{and}\quad
\epsilon_{0i}^{*}=Y_i-g_{0i}^{*}.
\]
Similar derivations to those used in the proof of Lemma \ref{lemma5}
show that, under the ``local'' alternatives
$\mathscr{L}_{{\beta}^{\ss2}}$, we have
\[
\widehat\bd
= \{\bi_n+(\bw^{*}_{20}-\bw^{*}_{10})\bb^{*}\}\bd^{*}
\]
where
\[
\bd^{*}=(\epsilon_{01}^{*}\psi_{01}^{*}, \ldots,
\epsilon_{0n}^{*}\psi_{0n}^{*})^{\T},\quad
\bb^{*}={\bx}^{\s1}\{\bi(\beta^{*\s1})\}^{-1}\bx^{\s1\T},\quad
\bi(\beta^{*\s1})={\bx}^{\s1\T}\bw^{*}_{10}\bx^{\s1},
\]
\[
\bw^{*}_{10}=\text{diag}\{\psi_{01}^{*2}E(\epsilon_{01}^{*2}|X_1),\ldots,\psi_{0n}^{*2}E(\epsilon_{0n}^{*2}|X_n)\}\quad
\text{and}\quad
\bw^{*}_{20}=\text{diag}\{\psi^{*'}_{01}g^{*'}_{01}\epsilon_{01}^{*},\ldots,\psi'_{0n}g^{*}_{0n}\epsilon_{0n}^{*}\}.
\]
Hence, \be \label{Atm50}
\begin{split}
\widetilde{U}_n=&
{n^{-1}}[{\bd}^{*\T}\{\bi_n+(\bw^{*}_{20}-\bw^{*}_{10})\bb^{*}\}^{\T}{\ba}\{\bi_n+(\bw^{*}_{20}-\bw^{*}_{10})\bb^{*}\}{\bd}^{*}]\\
=&{n^{-1}}\bd_1^{*\T}\ba^{*}\bd_1^{*}+{n^{-1}}{\bd}_2^{*\T}\ba^{*}\bd_2^{*}+{2}{n^{-1}}{\bd}_1^{*\T}\ba^{*}\bd^{*}_2=
T_1+T_2+T_3,\quad \text{say}, \end{split} \ee where $
\bd^{*}_1=(\epsilon_{1}\psi_{01}^{*}, \ldots,
\epsilon_{n}\psi_{0n}^{*})^{\T},~\bd_2^{*}=\{(g_1-g_{01}^{*})\psi_{01}^{*},
\ldots, (g_n-g_{0n}^{*})\psi_{0n}^{*}\}^{\T} $ and
\[
{\ba}^{*}=\{\bi_n+(\bw^{*}_{20}-\bw^{*}_{10}){\bb}^{*}\}^{\T}{\ba}\{\bi_n+(\bw^{*}_{20}-\bw^{*}_{10}){\bb}^{*}\}.
\]
Derivations  in \cite{Chen_Guo_2014} demonstrate that
\be \label{Atm52}
\frac{T_{1}}{[{2\text{tr}\{\Sigma^{\ss2}_{\beta}({\beta}_0^{\ast})\}^2}]^{1/2}}\to
N(0,1)\quad \mbox{in distribution as $n \to \infty$}. \ee \be
\label{Atm53}
T_{2}-n\Delta_{\beta,{\beta}_0^{\ast}}^{\ss2\T}\Delta_{\beta,{\beta}_0^{\ast}}^{\ss2}=o_p\bigl(\text{tr}^{{1}/{2}}[\{\Sigma^{\ss2}_{\beta}({\beta}_0^{\ast})\}^2]\bigr)\quad
\text{and}
\ee 
 \be
\label{Atm4}T_3=o_p\bigl(\text{tr}^{{1}/{2}}[\{\Sigma^{\ss2}_{\beta}({\beta}_0^{\ast})\}^2]\bigr).
\ee  Combining the results in (\ref{Atm50})-(\ref{Atm4}), under the
``local" alternatives $\mathscr{L}_{\beta^{\ss2}}$, we have \be
\frac{\widetilde{U}_n-n\Delta_{\beta,{\beta}_0^{\ast}}^{\ss2\T}\Delta_{\beta,{\beta}_0^{\ast}}^{\ss2}}{[{2\text{tr}\{\Sigma^{\ss2}_{\beta}({\beta}_0^{\ast})\}^2}]^{1/2}}\to
N(0,1)\nn \ee in distribution as $n\to \infty.$ \ov

\vskip 1cm

\noindent \textbf{Proof of Proposition \ref{tm4}}

\smallskip

\noindent
Derivations given in \cite{Chen_Guo_2014} show that \be \max_{1\leq
i \leq n}|\hat{\mu}_{0i}-\mu_{0i}|=o_p(1)\quad  \text{and}\quad
\max_{1\leq i \leq
n}|\hat{\psi}_{0i}^2-\psi^2_{0i}|=o_p(1).\label{prop_1}\ee

\noindent Note that
\be(Y_i-\hat{\mu}_{0i})^2=\epsilon_{0i}^2+(\hat{\mu}_{0i}-\mu_{0i})^2+2(\hat{\mu}_{0i}-\mu_{0i})\epsilon_{0i}\quad
\text{and}\quad
\hat{\psi}_{0i}^2=\psi_{0i}^2+\hat{\psi}_{0i}^2-\psi_{0i}^2.
\label{prop_2} \ee
 Thus we can write \[
\begin{split}
\widehat{R}_{n}
&=\frac{1}{n(n-1)}\sum\limits_{i \neq j
}^{n}\epsilon_{0i}^{2}\epsilon_{0j}^{2}\psi_{0i}^2\psi_{0j}^2(X_{i}^{\ss2\T}X_{j}^{\ss2})^{2}+\Delta
R_n= R_n+\Delta R_n,\quad \text{say}.
\end{split}
\]
$\Delta R_n$ is straightforward to obtain from (\ref{prop_2}) and
the definition of $\widehat{R}_{n}$ and hence is omitted here.
 Similar to the proofs in Lemma \ref{pn1}, we have
$R_n=\text{tr}\{\Sigma^{\ss2}_{\beta}({\beta}_0^{\ast})\}^2+o_p\bigl(\text{tr}[\{\Sigma^{\ss2}_{\beta}({\beta}_0^{\ast})\}^2]\bigr)$.
 Analogous to Lemma \ref{mean_variance_gt}, we
can show $\Delta
R_n=o_p\bigl(\text{tr}[\{\Sigma^{\ss2}_{\beta}({\beta}_0^{\ast})\}^2]\bigr)$
using (\ref{prop_1}). Hence we complete the proof of the
proposition. \ov

\vskip 1cm

\noindent \textbf{Proof of Corollary \ref{cy2}}

\smallskip

\noindent
Similar to the proofs in Proposition \ref{tm4}, under the ``local"
alternatives $\mathscr{L}_{{\beta}^{\ss2}}$,
\[\max_{1\leq i
\leq n}|\hat{\mu}_{0i}-\mu_{0i}^{*}|=o_p(1)~~\text{and}~~
\max_{1\leq i \leq n}|\hat{\psi}_{0i}^2-\psi_{0i}^{*2}|=o_p(1),
\]
where $\mu_{0i}^{*}=g(X_i^{\T}\beta_0^{*})$,
$\psi_0^{*}=\psi(X,\beta_0^{*})$ and
$\beta_0^{*}=(\beta^{*\s1\T},\beta_0^{\ss2\T})^{\T}$. Define
$\Sigma_0^{*}=E(\epsilon_0^{*2}\psi_0^{*2}X^{\ss2}X^{\ss2\T})$ where
$\epsilon_0^{*}=Y-\mu_0^{*}$. Employing the same technique as we
used in Proposition \ref{tm4},  \be \label{Atm54}
{\widehat{R}_n}/{\{\text{tr}(\Sigma_0^{*2})\}}\to 1\quad \mbox{in
probability as $n \to \infty$.} \ee
Under the ``local" alternatives $\mathscr{L}_{\beta^{\ss2}}$, it can
be shown that
\[
\begin{split}
\text{tr}(\Sigma_0^{*2})-\text{tr}[\{\Sigma^{\ss2}_{\beta}({\beta}_0^{\ast})\}^2]=o\bigl(\text{tr}[\{\Sigma^{\ss2}_{\beta}({\beta}_0^{\ast})\}^2]\bigr).
\end{split}
\]
This together with (\ref{Atm54}) implies  \[
\frac{\widehat{R}_n}{\text{tr}[\{\Sigma^{\ss2}_{\beta}({\beta}_0^{\ast})\}^2]}\to
1\quad \mbox{in probability as $n \to \infty$}.
\]
Hence, the power of the test is
\[
\begin{split}
\Omega^{\ss2}(\beta, {\beta}_0^{\ast})
=&\Phi\left(-z_{\alpha}+\frac{n\Delta_{\beta,{\beta}_0^{\ast}}^{\ss2\T}\Delta_{\beta,{\beta}_0^{\ast}}^{\ss2}}{\bigl[{2\text{tr}\{\Sigma^{\ss2}_{\beta}({\beta}_0^{\ast})\}^2}\bigr]^{1/2}}\right)\{1+o(1)\}.
\end{split}
\]
This completes the proof the corollary.
 \ov

\begin{figure}
\setlength{\abovecaptionskip}{0pt}
\setlength{\belowcaptionskip}{10pt} \captionstyle{center}
  \onelinecaptionsfalse  \caption{Empirical power profiles, for testing the global hypothesis,  of the proposed test (solid lines with triangles) and the test of \cite{Goeman_Houwelingen_Finos_2011_Biom} (dashed lines with circles). }
\includegraphics[width=1.0\textwidth]{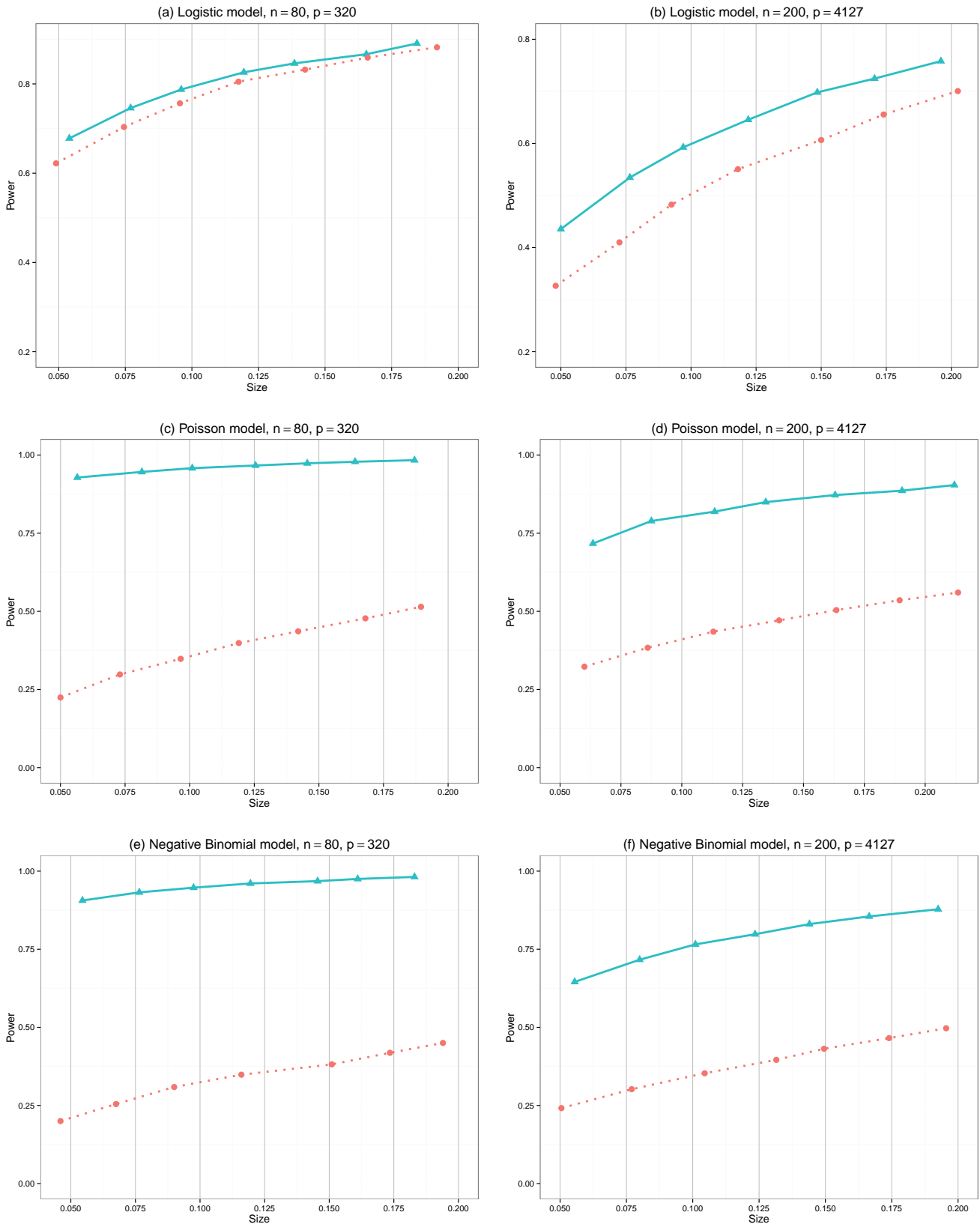}
\label{Figglobaltest}
\end{figure}

\begin{figure}
 \setlength{\abovecaptionskip}{0pt}
\setlength{\belowcaptionskip}{10pt} \captionstyle{center}
  \onelinecaptionsfalse  \caption{Empirical power profiles, for testing the hypothesis with nuisance parameters, of the proposed test (solid lines with triangles) and the test of \cite{Goeman_Houwelingen_Finos_2011_Biom} (dashed lines with circles).}
\includegraphics[width=1.0\textwidth]{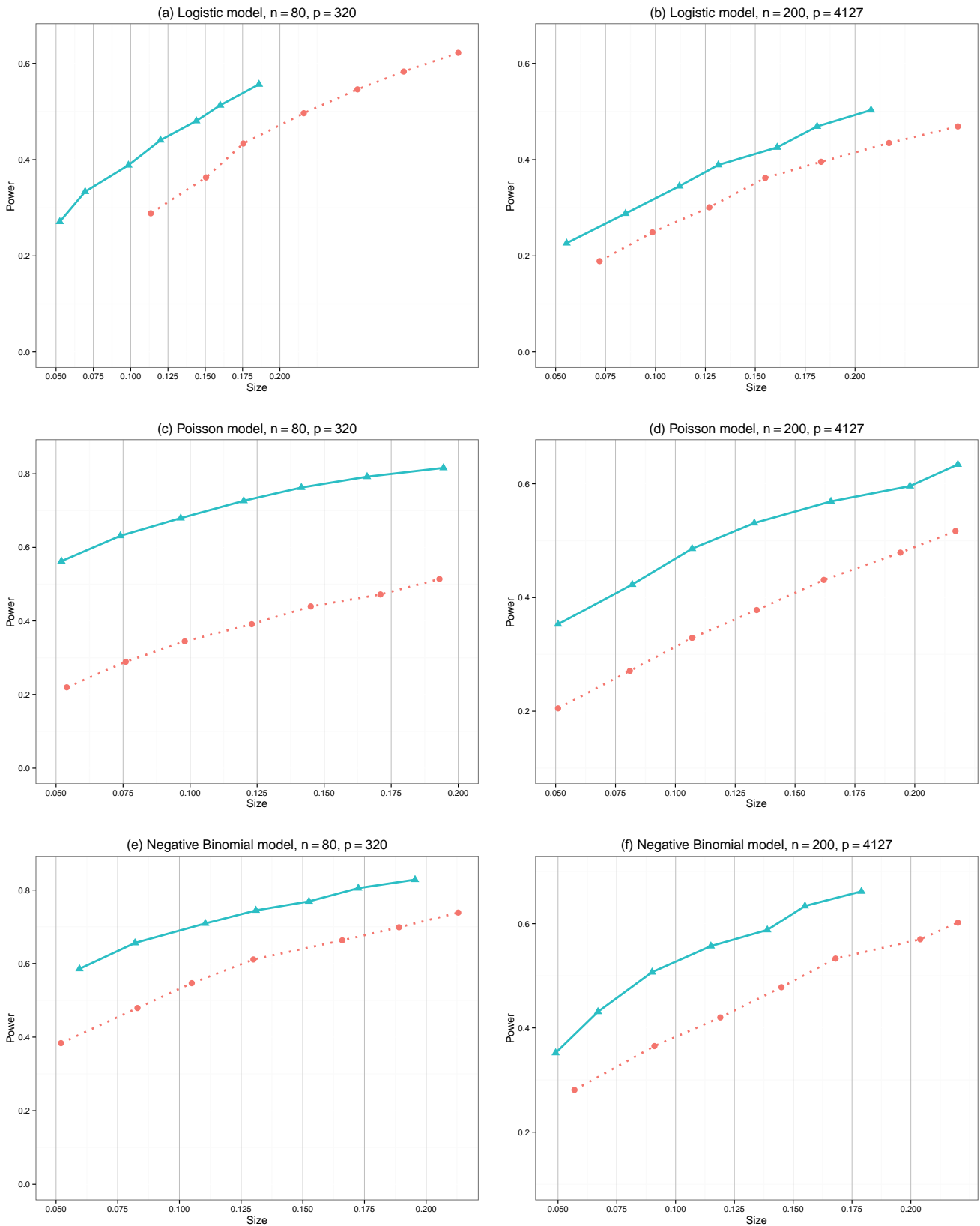}
\label{Fignuisance}
\end{figure}

\begin{figure}
 \setlength{\abovecaptionskip}{0pt}
\setlength{\belowcaptionskip}{10pt} \captionstyle{center}
\onelinecaptionsfalse \caption{Histograms of p-values (left panels)
and the standardized test statistic under the null hypothesis (right
panels) for the global hypothesis.}\centering
\includegraphics[scale=0.42]{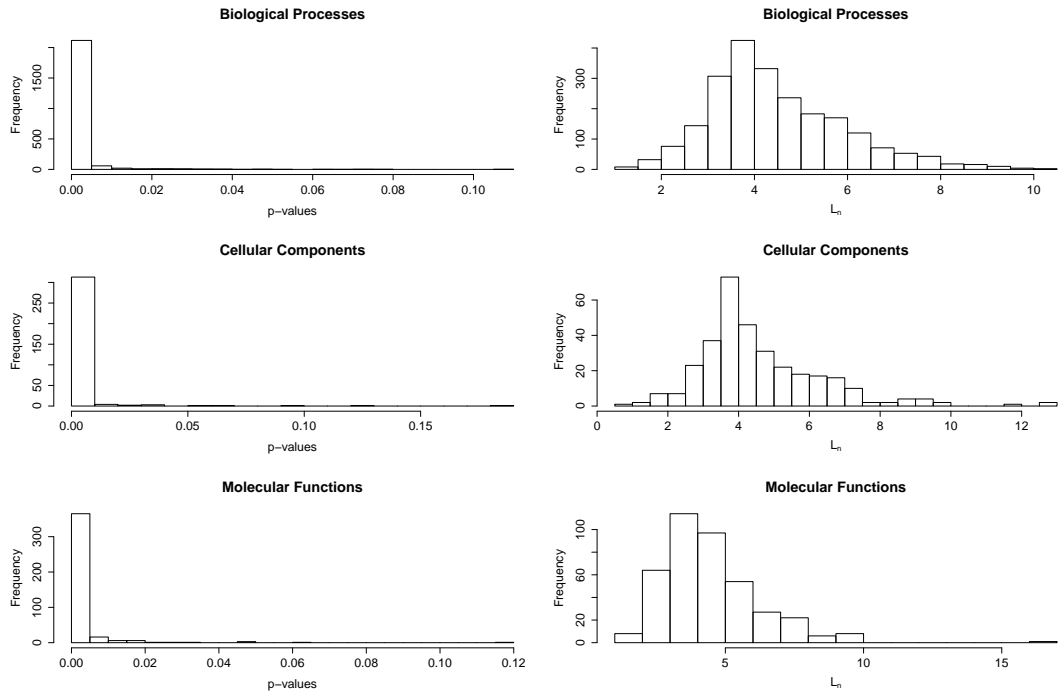}
\label{caseglobal}
\end{figure}

\begin{figure}
 \setlength{\abovecaptionskip}{0pt}
\setlength{\belowcaptionskip}{10pt} \captionstyle{center}
\onelinecaptionsfalse \caption{Histograms of p-values (left panels)
and the standardized test statistics (right panels) of the proposed
test  in the presence of  the nuisance parameter under the logistic
model.}\centering
\includegraphics[scale=0.42]{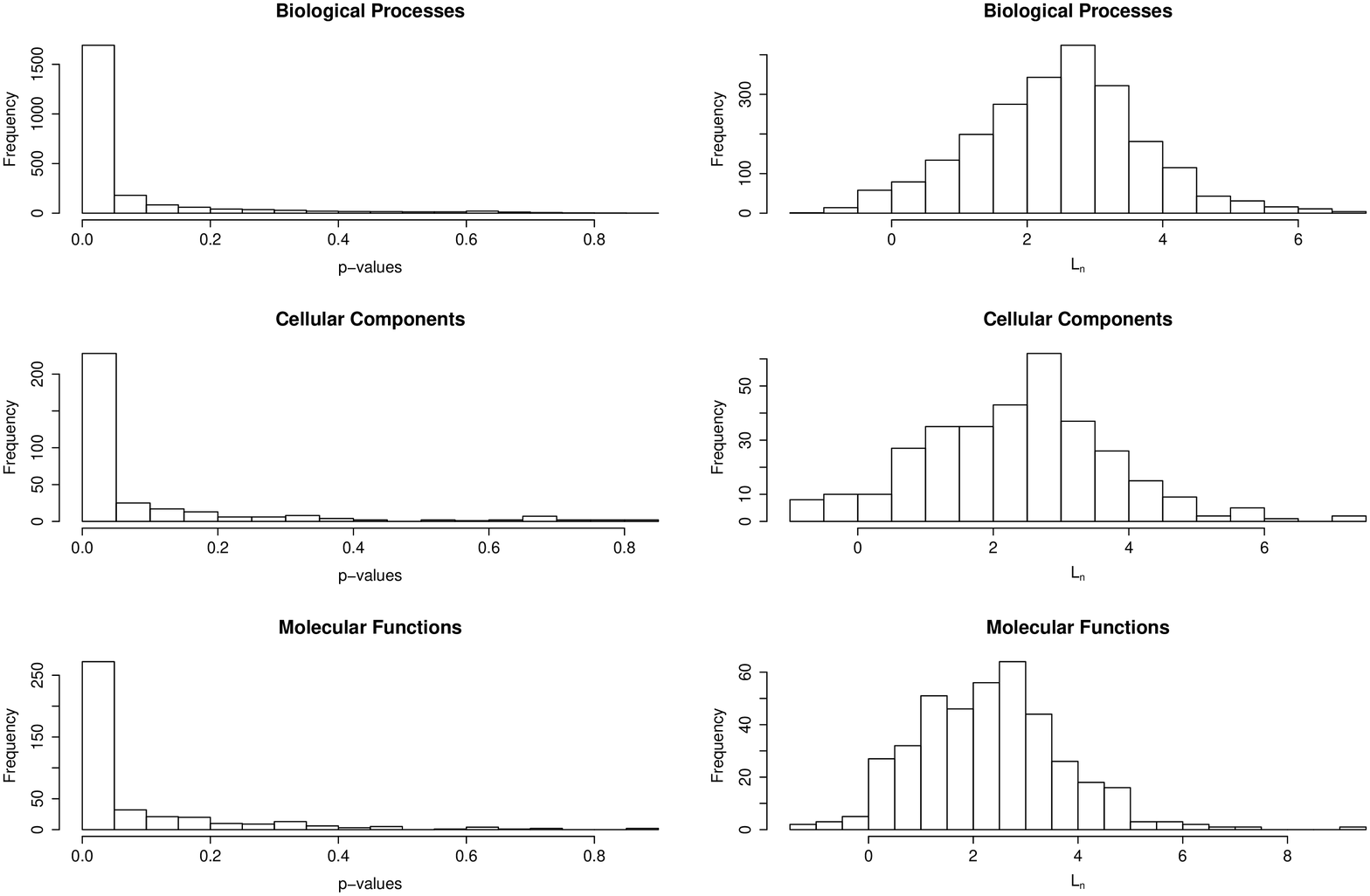}
\label{caselogiticnui}
\end{figure}

\begin{figure}
 \setlength{\abovecaptionskip}{0pt}
\setlength{\belowcaptionskip}{10pt} \captionstyle{center}
\onelinecaptionsfalse \caption{Histograms of p-values (left panels)
and the standardized test statistics (right panels) of the proposed
test in the presence of  the nuisance parameters under the probit
model.}\centering
\includegraphics[scale=0.42]{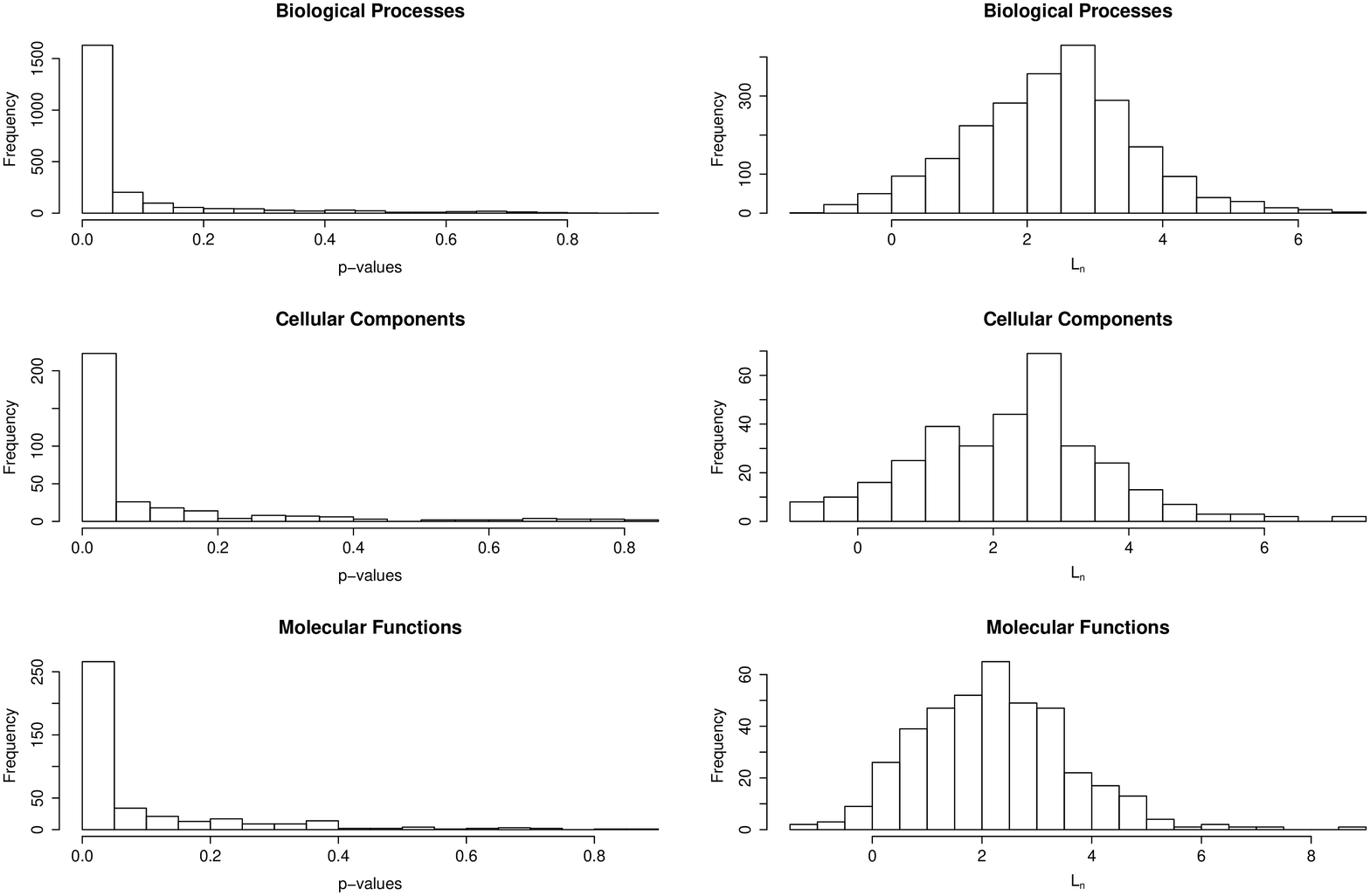}
\label{caseprobitnui}
\end{figure}

\begin{table}[bht]
\small
 \setlength{\abovecaptionskip}{0pt}
\setlength{\belowcaptionskip}{10pt}
\begin{center}
\captionstyle{center}
\caption{Two by two  classifications on the number (proportion) of
gene-sets   rejected/not rejected  in the tests with nuisance
parameter under the logistic and probit models.}

\begin{tabular}{cccccccccc}
\hline \hline
\multirow{2}{*}{} &  \multirow{2}{*}{} &  \multirow{2}{*}{} &  & \multicolumn{2}{c} {Probit model}   \\
 \cline{2-8}
Logistic model  & &  Rejected &  Not rejected &  Rejected &  Not rejected &  Rejected &  Not rejected\\
 \hline
 \multirow{2}{*}{} & \multirow{2}{*}{}  & \multicolumn{2}{c}{Biological Processes}   &  \multicolumn{2}{c}{Cellular Components}   &  \multicolumn{2}{c}{ Molecular Functions} \\
   Rejected       &       &   981(0.44)     & 103(0.05)       &   140(0.43)      & 14(0.04) & 113(0.28)      &  40(0.10)        \\
 Not rejected     &       &   0(0.00)      &  1161(0.51)        &    0(0.00)      &  174(0.53)  &   19(0.05)       & 230(0.57)  \\

 \hline
\hline
\end{tabular}
 \label{table2}
\end{center}
\end{table}


\begin{thebibliography}{11}

\newcommand{\enquote}[1]{``#1''}
\expandafter\ifx\csname
natexlab\endcsname\relax\def\natexlab#1{#1}\fi





\bibitem[{Arias-Castro et~al.(2011)}]{Arias_Candes_Plan_2011_AOS}
Arias-Castro, E., Cand\`es, E. J. and Plan, Y. (2011). {Global
testing under sparse alternatives: anova, multiple comparisons and
the higher criticism.} \textit{The Annals of Statistics},
\textbf{39}, 2533-2556.

\bibitem[{Auer and Doerge(2010)}]{Auer_Doerge_2010}
Auer, P. L. and Doerge, R. W. (2010). {Statistical design and
analysis of RNA sequencing data.} \textit{Genetics}, \textbf{185},
405-416.


\bibitem[{Bai and Saranadasa(1996)}]{Bai_Sara_1996_Sinica}
Bai, Z. D. and Saranadasa, H. (1996). {Effect of high dimension: by
an example of two sample problem.} \textit{Statistica Sinica},
\textbf{6}, 311-329.

\bibitem[{Benjamini and Hochberg(1995)}]{Benjamini_Hochberg_1995_JRSSB}
Benjamini, Y. and Hochberg, Y. (1995). {Controlling the false
discovery rate: A practical and powerful approach to multiple
testing.} \textit{Journal of the Royal Statistical Society: Series B
(Statistical Methodology)}, \textbf{57}, 289-300.





\bibitem[{Chang et al.(2013)}]{Chang_AOS}
Chang, J., Tang, C. Y. and Wu, Y. (2013). {Marginal empirical
likelihood and sure independence feature screening.} \textit{The
Annals of Statistics}, \textbf{41},  2123-2148.

\bibitem[{Chen and Guo(2014)}]{Chen_Guo_2014}
Chen, S. X. and Guo, B. (2014). { Tests for high dimensional
generalized linear models.} Technical report, Guanghua School of
Managment, Peking University.

\bibitem[{Chen et~al.(2009)}]{Chen_Peng_Qin_2009_Biom}
Chen, S. X., Peng, L. and Qin, Y. L. (2009). {Effects of data
dimension on empirical likelihood.} \textit{Biometrika},
\textbf{96}, 711-722.

\bibitem[{Chen and Qin(2010)}]{Chen_Qin_2010_AOS}
Chen, S. X. and Qin, Y. L. (2010). {A two-sample test for
high-dimensional data with applications to gene-set testing.}
\textit{The Annals of Statistics}, \textbf{38}, 808-835.


\bibitem[{Chen et~al.(2010)}]{Chen_Zhang_Zhong_2010_JASA}
Chen, S. X., Zhang, L. X. and Zhong, P. S. (2010). {Tests for
high-dimensional covariance matrices.} \textit{Journal of the
American Statistical Association}, \textbf{105}, 810-819.

\bibitem[{Chiaretti et~al.(2004)}]{Chiaretti_2004_Blood}
Chiaretti, S., Li, X., Gentleman, R., Vitale, A., Vignetti, M.,
Mandelli, F., Ritz, J. and Foa, R. (2004). {Gene expression profile
of adult T-cell acute lymphocytic leukemia identifies distinct
subsets of patients with different response to therapy and
survival.} \textit{Blood}, \textbf{103}, 2771-2778.





\bibitem[{Donoho and Jin(2004)}]{Donoho_Jin_2004}
Donoho, D. and Jin, J. (2004). {Higher criticism for detecting
sparse heterogeneous mixtures.} \textit{The Annals of Statistics},
\textbf{32}, 962-994.

\bibitem[{Dudoit et~al.(2008)}]{Dudoit_2008}
Dudoit, S., Keles, S. and van der Laan, M. J. (2008) {Multiple tests
of association with biological annotation metadata.}
\textit{Institute of Mathematical Statistics. Collections},
\textbf{2}, 153-218.


\bibitem[{Efron and Tibshirani(2007)}]{Efron_2007_AOAS}
Efron, B. and Tibshirani, R. (2007). {On testing the significance of
sets of genes.} \textit{The Annals of Applied Statistics},
\textbf{1}, 107-129.

\bibitem[{Fahrmeir and Tutz(1994)}]{FahrmeirTutz_1994}
Fahrmeir, L. and Tutz, G. (1994). \textit{%
 Multivariate statistical modelling based on generalized linear models} (2nd
 edition).
 Springer, New York.

\bibitem[{Fan and Song(2010)}]{Fan_Song_2010_AOS}
Fan, J. and Song, R. (2010). {Sure independent screening in
generalized linear models with NP-dimensionality.} \textit{The
Annals of Statistics}, \textbf{38}, 3567-3604.






\bibitem[{Fan and Lv(2008)}]{Fan_Lv_2008_JRSSB}
Fan, J. and Lv, J. (2008). {Sure independence screening for
ultrahigh dimensional feature space.} \textit{Journal of the Royal
Statistical Society: Series B (Statistical Methodology)},
\textbf{70}, 849-911.

\bibitem[{Gentleman et~al.(2005)}]{Gentleman_2005}
Gentleman, R., Irizarry, R. A., Carey, V. J., Dudoit, S. and Huber,
W. (2005).  \textit{%
Bioinformatics and Computational Biology Solutions Using R and
Bioconductor. } Springer, New York.


\bibitem[{Goeman et~al.(2006)}]{Goeman_Geer_Houwelingen_2006_JRSSB}
Goeman, J. J., Van De Geer, S. A. and Van Houwelingen, H. C. (2006).
{Testing against a high dimensional alternative.} \textit{Journal of
the Royal Statistical Society: Series B (Statistical Methodology)},
\textbf{68}, 477-493.

\bibitem[{Goeman et~al.(2011)}]{Goeman_Houwelingen_Finos_2011_Biom}
Goeman, J. J., Van Houwelingen, H. C. and Finos, L. (2011). {Testing
against a high-dimensional alternative in the generalized linear
model: asymptotic type I error control.} \textit{Biometrika},
\textbf{98}, 381-390.

\bibitem[{Hall and Heyde(1980)}]{Hall_Heyde_1980}
Hall, P., and Heyde, C. C. (1980). \textit{%
Martingale limit theory and its application.} Academic Press.


\bibitem[{Ingster et~al.(2010)}]{Ingster_Tsybakov_Verzelen_2010_EJS}
Ingster, Y. I., Tsybakov, A. B. and Verzelen, N. (2010). {Detection
boundary in sparse regression.} \textit{Electronic Journal of
Statistics}, \textbf{4}, 1476-1526.


\bibitem[{Lan et~al.(2014)}]{Lan_Wang_Tsai_2013}
Lan, W., Wang, H. and Tsai, C. L. (2014). {Testing covariates in
high-dimensional regression.} \textit{Annals of Institute of
Statistical Mathematics}, DOI:
10.1007/s10463-013-0414-0.


\bibitem[{Le Cessie and Van Houwelingen(1991)}]{le Cessie_van Houwelingen_1991_Bio}
Le Cessie, S. and Van Houwelingen, J. C. (1991). {A goodness-of-fit
test for binary regression models, based on smoothing methods.}
\textit{Biometrics}, \textbf{47}, 1267-1282.



\bibitem[{Li and Chen(2012)}]{Li_Chen_2012_AOS}
Li, J. and Chen, S. X. (2012). {Two sample tests for
high-dimensional covariance matrices.} \textit{The Annals of
Statistics}, \textbf{40}, 908-940.



\bibitem[{Lund et~al.(2012)}]{Lund_2012}
Lund, S., Nettleton, D., McCarthy, D. and Smyth, G. (2012). {
Detecting differential expression in RNA-sequence data using
quasi-likelihood with shrunken dispersion estimates.}
\textit{Statistical applications in genetics and molecular biology},
\textbf{11}, 8.



\bibitem[{McCullagh(1983)}]{McCullagh_1983_AOS}
McCullagh, P. (1983). {Quasi-likelihood functions.} \textit{The
Annals of Statistics}, \textbf{11}, 59-67.

\bibitem[{McCullagh and Nelder(1989)}]{McCullagh_Nelder_(1989)}
McCullagh, P. and Nelder, J. A. (1989). \textit{Generalized Linear
models} (2nd edition). Chapman and Hall.






\bibitem[{Pan(2009)}]{Pan_2009}
Pan, W. (2009).  {Asymptotic tests of association with multiple SNPs
in linkage disequilibrium.} \textit{Genetic epidemiology},
\textbf{33}, 497-507.


\bibitem[{Rahmatallah et~al.(2012)}]{Rahmatallah_2012_Bioinformatics}
Rahmatallah, Y., Emmert-Streib, F. and Glazko, G. (2012). {Gene set
analysis for self-contained tests: complex null and specific
alternative hypotheses.} \textit{Bioinformatics}, \textbf{28},
3073-3080.

\bibitem[{Seber(2008)}]{Seber_2008}
 Seber G. A. (2008). \textit{%
A matrix handbook for statisticians}, Wiley, New York.

\bibitem[{Serfling(1980)}]{Serfling_1980}
Serfling, R. J. (1980). \textit{Approximate theorems of mathematical
statistics.} Wiley, New York.













\bibitem[{van de Geer(2008)}]{Geer_2006_AOS}
van de Geer, S. (2008).  {High-dimensional generalized linear models
and the lasso.} \textit{The Annals of Statistics}, \textbf{36},
614-645.

\bibitem[{Wedderburn(1974)}]{Wedderburn_1974_Biom}
Wedderburn, R. W. (1974).  {Quasi-likelihood functions, generalized
linear models, and the Gauss-Newton method.} \textit{Biometrika},
\textbf{61}, 439-447.

\bibitem[{White(1982)}]{White_1982_Eco}
White, H. (1982).  {Maximum likelihood estimation of misspecified
models.} \textit{Econometrica}, \textbf{50}, 1-25.




\bibitem[{Zhong and Chen(2011)}]{Zhong_Chen_2011_JASA}
Zhong, P. S. and Chen, S. X. (2011).  {Tests for high dimensional
regression coefficients with factorial designs.} \textit{Journal of
the American Statistical Association,} \textbf{106}, 260-274.
\end{thebibliography}
\end{document}